\documentclass[english,journal, twocolumn, a4paper, 10pt]{IEEEtran}
\usepackage[T1]{fontenc}
\usepackage[latin9]{inputenc}
\usepackage{color}
\usepackage{float}
\usepackage{bm}
\usepackage{amsmath}
\usepackage{amsthm}
\usepackage{amssymb}
\usepackage{graphicx}

\makeatletter

\floatstyle{ruled}
\newfloat{algorithm}{tbp}{loa}
\providecommand{\algorithmname}{Algorithm}
\floatname{algorithm}{\protect\algorithmname}

\theoremstyle{plain}
\newtheorem{thm}{\protect\theoremname}
\theoremstyle{plain}
\newtheorem{lem}[thm]{\protect\lemmaname}

\usepackage{bm}
\usepackage{subfigure}
\usepackage{cite}
\usepackage[english]{babel}
\usepackage[T1]{fontenc}
\usepackage{algorithm}
\usepackage{algorithmic}
\makeatletter
\def\blfootnote{\xdef\@thefnmark{}\@footnotetext}
\makeatother

\makeatother

\usepackage{babel}
\providecommand{\lemmaname}{Lemma}
\providecommand{\theoremname}{Theorem}

\begin{document}

\title{Cache Placement in Fog-RANs: \\From Centralized to Distributed Algorithms\thanks{This work was supported in part by the NSFC under Grant No. 61601255,
the Hong Kong Research Grants Council under Grant No. 610113, the
Scientific Research Foundation of Ningbo University under Grant No.
010-421703900 and the Zhejiang Open Foundation of the Most Important
Subjects under Grant No. 010-421500212. This work was presented in
part at the IEEE International Conference on Communications (ICC),
Kuala Lumpur, Malaysia, May 2016\cite{2016_Jliu_caching_ICC}. \newline J.
Liu is with the College of Electrical Engineering and Computer Science,
Ningbo University, Zhejiang, China, 315211. E-mail: eeliujuan@gmail.com.\newline B.
Bai is the Future Network Theory Lab, Huawei Technologies Co., Ltd.,
Shatin, N. T., Hong Kong. E-mail: baibo8@huawei.com. \newline J. Zhang
and K. B. Letaief are with the Department of Electronic and Computer
Engineering, The Hong Kong University of Science and Technology, Clear
Water Bay, Hong Kong. K. B. Letaief is also with Hamad bin Khalifa
University, Doha, Qatar. E-mail:eejzhang@ust.hk, eekhaled@ust.hk. } }
\author{Juan Liu, \IEEEmembership{Member, IEEE}, Bo Bai, \IEEEmembership{Member, IEEE},
Jun Zhang, \IEEEmembership{Senior Member, IEEE}, \\and Khaled B. Letaief, \IEEEmembership{Fellow, IEEE}}

\maketitle

\begin{abstract}
To deal with the rapid growth of high-speed and/or ultra-low latency
data traffic for massive mobile users, fog radio access networks (Fog-RANs)
have emerged as a promising architecture for next-generation wireless
networks. In Fog-RANs, the edge nodes and user terminals possess storage,
computation and communication functionalities to various degrees,
which provides high flexibility for network operation, i.e., from
fully centralized to fully distributed operation. In this paper, we
study the cache placement problem in Fog-RANs, by taking into account
flexible physical-layer transmission schemes and diverse content preferences
of different users. We  develop both centralized and distributed transmission
aware cache placement strategies to minimize users\textquoteright{}
average download delay subject to the storage capacity constraints.
In the centralized mode, the cache placement problem is transformed
into a matroid constrained submodular maximization problem, and an
approximation algorithm is proposed to find a solution within a constant
factor to the optimum. In the distributed mode, a belief propagation
based distributed algorithm is proposed to provide a suboptimal solution,
with iterative updates at each BS based on locally collected information.
Simulation results  show that by exploiting caching and cooperation
gains, the proposed transmission aware caching algorithms can greatly
reduce the users\textquoteright{} average download delay. 
\end{abstract}


\section{Introduction}

With the explosive growth of consumer-oriented multimedia applications,
a large scale of end devices, such as smart phones, wearable devices
and vehicles, need to be connected via wireless networking  \cite{mung_chiang_fog_2016}.
This has triggered the rapid increase of high-speed and/or ultra-low
latency data traffic that is very likely generated, processed and
consumed locally at the edge of wireless networks. To cope with this
trend, fog radio access network (Fog-RAN) is emerging as a promising
network architecture, in which the storage, computation, and communication
functionalities are moved to the edge of wireless networks, i.e.,
to the near-user edge devices and end-user terminals \cite{mung_chiang_fog_2016,seok-hwan_park_joint_2016,shi_large-scale_2015}.
To further improve the delivery rate and decrease latency for mobile
users, a promising solution is to push the popular contents towards
end users by caching them at the edge nodes in Fog-RANs \cite{seok-hwan_park_joint_2016}.
Thus, the content delivery service of mobile users consists of two
phases, i.e., \emph{cache placement} and \emph{content delivery} \cite{2016_Jliu_caching_ICC,borst_distributed_2010,golrezaei_base-station_2012,maddah-ali_fundamental_2014,golrezaei_femtocaching:_2012,2015_XP_caching}.
The recent works studying cache-aided wireless networks fall into
two major categories: 1) analyzing the content delivery performance
for certain cache placement policies; 2) designing cache placement
strategies for efficient content delivery. 

It is critical to study the content delivery performance in cache-assisted
wireless networks to reveal the benefits of placing caches distributedly
across the whole network \cite{2015bacstuvgcacheenabled,ji_optimal_2013,ji_wireless_2016,jeon_wireless_2015,liu_improvement_2015,2016_MTao_TWC,2016_MTao_TWC_multicast}.
By coupling physical-layer transmission and random caching, the authors
in \cite{2015bacstuvgcacheenabled} investigated the system performance
in terms of the average delivery rate and outage probability for small-cell
networks, where cache-enabled BSs are modeled as a Poisson point process.
In \cite{ji_optimal_2013} and \cite{ji_wireless_2016}, the throughput-outage
tradeoff was investigated and the throughput-outage scaling laws were
revealed for cache-assisted wireless networks, where clustered device
caching and one-hop device-to-device (D2D) transmission are applied.
 This line of works have also been extended to the multi-hop D2D
network in \cite{jeon_wireless_2015}, where the multi-hop capacity
scaling laws were studied. The throughput scaling laws were studied
for wireless Ad-Hoc networks with device caching in \cite{liu_improvement_2015},
where the maximum distance separable (MDS) code and cache-assisted
multi-hop transmission/cache-induced coordinate multipoint (CoMP)
delivery were applied. In \cite{2016_MTao_TWC} and \cite{2016_MTao_TWC_multicast},
content-centric multicasting was studied for cache-enabled cloud RAN
and heterogeneous cellular networks, respectively. 

Cache placement strategies should be carefully designed such that
flexible transmission opportunities can be provided among users and
caching gain can be efficiently exploited in the content delivery
phase \cite{2016_Jliu_caching_ICC,golrezaei_femtocaching:_2012,niesen_coded_2014,maddah-ali_fundamental_2014,maddah-ali_decentralized_2015,karamchandani_hierarchical_2014,ahlehagh_video-aware_2014,bastug_SPAWC_2015,2015_XP_caching,2016_RWang_ComMag,poularakis_approximation_2014,2015_MDehghan_Infocom}.
The cache placement problem in femtocell networks was studied in \cite{golrezaei_femtocaching:_2012},
where femtocell BSs with finite-capacity storages are deployed to
act as helper nodes to cache popular files. In \cite{niesen_coded_2014,maddah-ali_fundamental_2014},
coded caching was exploited to create simultaneous coded multicasting
opportunities to mobile users. This work was extended to the decentralized
setting in \cite{maddah-ali_decentralized_2015} and  hierarchical
two-layer network in \cite{karamchandani_hierarchical_2014}, respectively.
By applying an Alternating Direction Method of Multipliers approach,
the authors of \cite{bastug_SPAWC_2015} proposed a distributed caching
algorithm for cache-enabled small base stations (SBSs) to minimize
the global backhaul costs of all the SBSs subject to the cache storage
capacities. In \cite{2015_XP_caching}, the design of optimal cache
placement was pursued for wireless networks, by taking the extra delay
induced via backhaul links and physical-layer transmissions into consideration.
The authors in \cite{ahlehagh_video-aware_2014} proposed user preference
profile based caching policies for radio access networks along with
backhaul and wireless channel scheduler to support more concurrent
video sessions. In \cite{2016_RWang_ComMag}, mobility-aware caching
strategies were proposed to exploit user mobility patterns to improve
cache performance. The joint routing and caching problem was studied
for small-cell networks and heterogeneous networks in \cite{poularakis_approximation_2014}
and \cite{2015_MDehghan_Infocom}, respectively, subject to both the
storage and transmission bandwidth capacity constraints on the small-cell
BSs.

The existing works mainly focused on designing centralized cache placement
strategies for specific network structures (e.g. small cell networks),
where some specific transmission schemes are applied for content delivery.
However, very few works have studied the cache placement problem in
Fog-RANs. We notice that different users may be connected to Fog-RANs
in different ways and with different transmission opportunities. Meanwhile,
Fog-RANs support flexible network operation, i.e., from fully centralized
to fully distributed operation. This motivates us to develop both
centralized and distributed transmission aware cache placement strategies
for the emerging Fog-RANs so that the spectrum efficiency of content
delivery is improved as much as possible. 

In this paper, we consider a Fog-RAN system, where each user is served
by one or multiple network edge devices, e.g., base stations (BSs),
and each BS is equipped with a cache of finite capacity.  In contrast
to \cite{golrezaei_femtocaching:_2012} and \cite{2015_MDehghan_Infocom}
where each user has the same file preference and file delivery scheme,
we consider that the users have different file preferences \cite{breslau_web_1999}
and possibly different candidate transmission schemes. Then, we formulate
an optimization problem to minimize the users' average download delay
subject to the BSs\textquoteright{} storage capacities, which turns
out to be NP-hard. To deal with this difficulty, we apply different
optimization techniques to find efficient cache placement policies
for centralized and distributed operation modes of Fog-RANs, respectively. 

In the centralized mode, we transform the delay minimization problem
into a matroid constrained submodular maximization problem \cite{2003_AlexanderSchrijver_combopt}.
In this problem, the average delay function is submodular for all
the possible transmission schemes, and the cache placement strategy
subject to the BSs\textquoteright{} storage capacities is a partition
matroid. Based on the submodular optimization theory \cite{2003_AlexanderSchrijver_combopt},
we then develop a centralized low-complexity algorithm to find a caching
solution within $1/2$ of the optimum in polynomial-time complexity
$\mathcal{O}(MNK)$, where $M$, $N$ and $K$ denote the number of
BSs, files and users, respectively. 

In the distributed mode, we develop a low-complexity belief propagation
based distributed algorithm to find a suboptimal cache placement strategy
\cite{kschischang_factor_2001}. Based on local information of its
storage capacity, the users in its serving range and their file request
statistics, each BS  perform individual computation and exchange its
belief on the local caching strategy with its neighboring BSs iteratively.
Through iterations, the distributed algorithm converges to a suboptimal
caching solution which achieves an average delay performance comparable
to the centralized algorithm, as shown by simulation results. By distributing
computing tasks, each individual BS always does much fewer calculations
than the central controller when running the caching algorithms. Notice
that the distributed caching algorithm proposed in \cite{bastug_SPAWC_2015}
is run by each SBS individually and no parameters are shared between
the SBSs. In this work, we propose a belief propagation based transmission
aware distributed caching algorithm which requires cooperation and
message passing between neighboring BSs. 

The rest of this paper is organized as follows. Section \ref{sec:System-Model}
introduces the system model of Fog-RANs. Section \ref{sec:Problem-Formulation}
formulates the cache placement problem which minimizes the average
download delay under the cache capacity constraints. In Section \ref{sec:Submodular-Optimization},
a centralized algorithm is proposed to solve the cache placement problem
under the framework of submodular optimization for the centralized
Fog-RANs. In Section \ref{sec:Distributed-Caching}, a belief propagation
based distributed algorithm is proposed for cache placement in the
distributed Fog-RANs.  Section \ref{sec:Simulation-Results} demonstrates
the simulation results. Finally, Section \ref{sec:Conclusions} concludes
this paper.

\section{System Model \label{sec:System-Model}}

\begin{figure}[t]
\centering
\renewcommand{\figurename}{Fig.}

\includegraphics[width=0.5\textwidth]{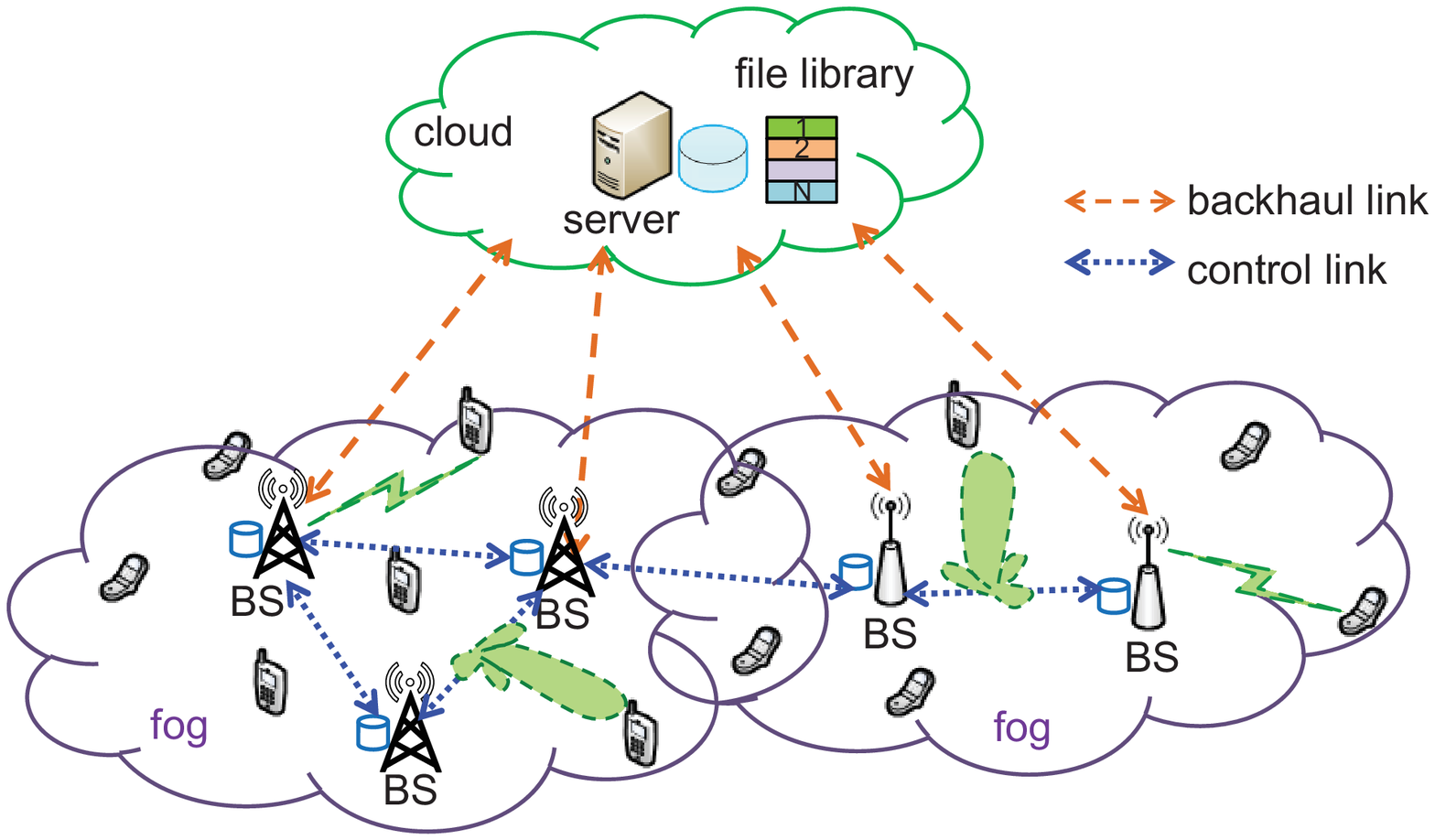}\caption{\textcolor{black}{An illustration of a Fog-RAN that consists of BSs
and mobile users, where BSs are connected to a cloud data center via
backhaul links. With the aid of transmission aware caching designs,
the neighboring BSs could cache the same files and deliver them to
their common users via cooperative beamforming. }}
\label{fig:system_model}
\end{figure}
As shown in Fig.$\,$\ref{fig:system_model}, we consider a Fog-RAN
consisting of $M$ edge nodes, i.e., BSs, and $K$ mobile users. Let
$\mathcal{A}=\left\{ a_{1},\cdots,a_{M}\right\} $ and $\mathcal{U}=\left\{ u_{1},\cdots,u_{K}\right\} $
denote the BS set and the user set, respectively. Each user can be
served by one or multiple BSs, depending on the way it connects to
the Fog-RAN. The connectivity between the users and the BSs is denoted
by a $K\times M$ matrix $\bm{L}$, where each binary element $l_{km}$
indicates whether user $u_{k}$ can be served by BS $a_{m}$. That
is, $l_{km}=1$ if user $u_{k}$ is located in the coverage of BS
$a_{m}$, and $l_{km}=0$ otherwise. The set of users in the coverage
of  BS $a_{m}$ is denoted by $\mathcal{U}_{m}=\{u_{k}\in\mathcal{U}|l_{km}=1\}$.
Similarly, the set of serving BSs of user $u_{k}$ is denoted by $\mathcal{A}_{k}=\{a_{m}\in\mathcal{\mathcal{A}}|l_{km}=1\}$. 

Suppose that the library of $N$ files, denoted by $\mathcal{F}=\left\{ f_{1},\cdots,f_{N}\right\} $,
is stored at one or multiple content servers which could be far away
in the cloud data center. The content servers can be accessed by the
BSs via backhaul links, as illustrated in Fig.$\,$\ref{fig:system_model}.
Assume all the files have the same size, i.e., $|f_{n}|=|f|$ $(\forall f_{n}\in\mathcal{F})$.
The file popularity distribution conditioned on the event that user
$u_{k}$ makes a request is denoted by $p_{nk}$, which can be viewed
as the user preference indicator and estimated via some learning procedure
\cite{bastug_transfer_2015,bharath_learning-based_2015}. The user's
file preferences are normalized such that $\sum_{n=1}^{N}p_{nk}=1$.
We also assume that each BS $a_{m}$ has a finite-capacity storage.
Denote by $Q_{m}$ the normalized storage capacity of BS $a_{m}$,
which means that each BS $a_{m}$ can store at most $Q_{m}$ files.
Let $x_{nm}$ be a binary variable indicating whether file $f_{n}$
is cached at BS $a_{m}$. That is, $x_{nm}=1$ if file $f_{n}$ is
stored at BS $a_{m}$, and otherwise $x_{nm}=0$. The caching variables
$\{x_{nm}\}$ shall be determined collaboratively by the BSs to improve
the probability that the users' requested files can be found in the
caches of the BSs, i.e., the hit probability.  Meanwhile, the cooperative
caching strategy, denoted by $\bm{X}$, should also be carefully designed
to provide flexible and cooperative transmission opportunities for
each user.

When user $u_{k}$ makes a request for file $f_{n}$, the serving
BSs $\mathcal{A}_{k}$  jointly decide how to transmit to this user
based on the caching strategy $\bm{X}$. Specifically, when file $f_{n}$
is cached in one or multiple BSs, the BSs  transmit this file to the
user directly by employing some transmission schemes, e.g., non-cooperative
transmission or cooperative beamforming, as shown in Fig.$\,$\ref{fig:system_model}.
When file $f_{n}$ has not been cached in any serving BS of the user,
the associated BSs $\mathcal{A}_{k}$  fetch the file from a content
server via backhaul links before they transmit to user $u_{k}$ over
wireless channels. 

The users' file delivery performance depends not only on the cache
placement strategy but also on the specific transmission schemes applied
to deliver the files to the users. In the following, we discuss the
file delivery rates for some typical physical-layer transmission schemes,
when the requested file is cached in one or multiple associated BSs.

\subsubsection{Non-cooperative Transmission}

When user $u_{k}$ is served by one single BS $a_{m}$, a non-cooperative
transmission scheme  be applied by this BS to transmit the file to
the user directly, if the requested file $f_{n}$ is cached in this
BS. Assume that efficient interference management schemes are applied
and interference power is constrained by a fixed value $\chi$. Let
$\mathrm{SINR}_{m}=\frac{P_{m}}{N_{0}B+\chi}$ denote the target signal-to-interference-plus-noise
ratio (SINR) at the transmitter side, where $P_{m}$ is the average
transmission power at BS $a_{m}$, $N_{0}$ is the power spectral
density of noise, and $B$ is the system bandwidth. The file delivery
rate in time slot $i$ can be estimated as 
\begin{equation}
R_{nk}(\bm{X},i)=B\log\left(1+\left|h_{km}(i)\right|^{2}l_{km}x_{nm}\mathrm{SINR}_{m}\right),\label{eq:fixedService}
\end{equation}
where $h_{km}(i)$ denotes the channel coefficient between user $u_{k}$
and BS $a_{m}$ in time slot $i$. 

\subsubsection{Cooperative Beamforming}

When user $u_{k}$ is served by multiple BSs, cooperative beamforming
can be applied by the associated BSs $\mathcal{A}_{k}$, if file $f_{n}$
has been cached in multiple BSs and the instantaneous channel state
information is available. During the file delivery phase, cooperative
beamformer can be created possibly in a distributed way to avoid signaling
overhead \cite{2007_RMudumbai_beamforming}. Accordingly, the file
delivery rate in time slot $i$ is estimated as 
\begin{equation}
R_{nk}(\bm{X},i)=B\log\left(1+\sum_{a_{m}\in\mathcal{A}_{k,n}}\left|h_{km}(i)\right|^{2}x_{nm}\mathrm{SINR}_{m}\right),\label{eq:Beamforming}
\end{equation}
where $\mathcal{A}_{k,n}\subseteq\mathcal{A}_{k}$ denotes a set of
BSs that transmit file $f_{n}$ to user $u_{k}$ via cooperative beamforming.

In this work, we aim at finding the optimal cache placement strategy
to minimize the average download delay, considering different candidate
transmission schemes for each user, as  be presented in the next section.

\section{Problem Formulation for Cache Placement\label{sec:Problem-Formulation}}

In this section, we first show how to calculate the average download
delay by applying martingale theory \cite{david_williams_probability_1991}.
Then, we formulate the cache placement problem. 

Let $\bar{D}_{nk}(\bm{X})$ denote the average delay for user $u_{k}$
to download file $f_{n}$ from its serving BSs for a given caching
strategy $\bm{X}$ and a specific transmission scheme. When file $f_{n}$
has been cached in one or multiple BSs,  user $u_{k}$ can download
this file from the associated BSs with rate $R_{nk}(\bm{X},i)$ (c.f.
(\ref{eq:fixedService})-(\ref{eq:Beamforming})) in each time slot
$i$. In this case, it takes at least $T_{nk}^{*}(\bm{X})$ time slots
for user $u_{k}$ to successfully receive all the bits of file $f_{n}$.
The minimum number of time slots $T_{nk}^{*}(\bm{X})$ can be evaluated
as
\begin{equation}
T_{nk}^{*}(\bm{X})=\arg\min\left\{ T:\sum_{i=1}^{T}R_{nk}(\bm{X},i)\geq\frac{\left|f_{n}\right|}{\Delta t}\right\} ,\label{eq:StoppingTime}
\end{equation}
where $\Delta t$ is the duration of one time slot. Thus, for user
$u_{k}$, the average delay of downloading file $f_{n}$ is expressed
as
\begin{equation}
\bar{D}_{nk}(\bm{X})=\mathbb{E}_{\bm{h}}\left\{ T_{nk}^{*}(\bm{X})\right\} \Delta t.
\end{equation}
When file $f_{n}$ has not been cached at any associated BS, one or
multiple serving BSs of user $u_{k}$, denoted by $\mathcal{A}_{k}^{'}$,
should first fetch the file from the content server via the backhaul
link before delivering the requested file to this user over wireless
channel. Let $D_{nk}$ denote the extra delay of downloading file
$f_{n}$ from the content server to the BSs $\mathcal{A}_{k}^{'}$.
We then evaluate the average download delay under the assumption that
the channel coefficients $\{h_{km}(i)\}$ are identically and independently
distributed (i.i.d.) across the time slots $i$ in the following theorem.

\begin{thm}
\label{thm:avdelay}If the channel coefficients $\{h_{km}(i)\}$ are
i.i.d. across the time slots, the average delay for user $u_{k}$
to download file $f_{n}$ can be expressed as 
\begin{equation}
\bar{D}_{nk}(\bm{X})=\begin{cases}
\frac{|f_{n}|}{\mathbb{E}_{\bm{h}}\left\{ R_{nk}(\bm{X})\right\} }, & \sum_{a_{m}\in\mathcal{A}_{k}}x_{nm}\neq0,\\
D_{nk}+\frac{|f_{n}|}{\mathbb{E}_{\bm{h}}\left\{ R_{nk}(\bm{X}_{k})\right\} }, & \sum_{a_{m}\in\mathcal{A}_{k}}x_{nm}=0.
\end{cases}\label{eq:av_Delay}
\end{equation}
 where $\mathbb{E}_{\bm{h}}\left\{ \cdot\right\} $ denotes the expectation
over the channel coefficients $\{h_{km}(i)\}$ and $\bm{X}_{k}$ is
a caching strategy with $x_{nm}=1$ for $a_{m}\in\mathcal{A}_{k}^{'}$.
 
\end{thm}
\begin{IEEEproof}
The proof is deferred to Appendix \ref{subsec:Theorem1_delay}.
\end{IEEEproof}
From this theorem, we can evaluate the average download delay by (\ref{eq:av_Delay})
for any given caching strategy and employed transmission scheme. Without
loss of generality, we assume that the users' average delay of downloading
file $f_{n}$ from the content server is larger than the average delay
of direct file delivery from the BSs and the following inequality
holds:
\begin{equation}
\frac{|f_{n}|}{\mathbb{E}_{\bm{h}}\left\{ R_{nk}(\bm{X}_{k})\right\} }+D_{nk}>\max_{\sum_{a_{m}\in\mathcal{A}_{k}}x_{nm}\neq0}\left\{ \frac{|f_{n}|}{\mathbb{E}_{\bm{h}}\left\{ R_{nk}(\bm{X})\right\} }\right\} .
\end{equation}
If $D_{nk}$ is much larger than $\frac{|f_{n}|}{\mathbb{E}_{\bm{h}}\left\{ R_{nk}(\bm{X}_{k})\right\} }$,
the average delay $\bar{D}_{nk}(\bm{X})$ can be approximated by $D_{nk}$
when $\sum_{a_{m}\in\mathcal{A}_{k}}x_{nm}=0$. Notice that $D_{nk}$
is the sum of the delay of file delivery within the Internet which
mainly depends on the level of congestion in the network, and the
delay of file delivery via backhaul links which may depend on the
backhaul capacities and the caching strategy $\bm{X}$. Considering
all these effects, the impact of the caching strategy $\bm{X}$ on
the delay $D_{nk}$ is negligible. Hence, we assume that the average
delay $D_{nk}$ is fixed and can be evaluated by the average time
of downloading file $f_{n}$ from the content server to the serving
BSs of user $u_{k}$. 

  In the considered system, we seek to design transmission aware
cache placement strategies to minimize the average delay of all the
users, by  taking different candidate transmission schemes for each
user into consideration.   Formally, the cache placement problem
can be formulated as follows 
\begin{equation}
\begin{split}\underset{\{x_{nm}\}}{\text{minimize}} & \quad\bar{D}(\bm{X})=\frac{1}{K}\sum_{k=1}^{K}\sum_{n=1}^{N}p_{nk}\bar{D}_{nk}(\bm{X})\\
\text{subject to} & \quad\begin{cases}
\sum_{n=1}^{N}x_{nm}\leq Q_{m},\forall a_{m}\in\mathcal{A}, & (a)\\
x_{nm}\in\{0,1\},\forall f_{n}\in\mathcal{F},a_{m}\in\mathcal{A}, & (b)
\end{cases}
\end{split}
\label{eq:opt_problem}
\end{equation}
where constraint ($\ref{eq:opt_problem}$.a) means that each BS $a_{m}$
is allowed to store at most $Q_{m}$ files. Since the variable $x_{nm}$
is binary, Problem $\eqref{eq:opt_problem}$ is a constrained integer
programming problem, which is generally NP-hard \cite{shanmugam_femtocaching:_2013}.
Hence, it is very challenging to find the optimal solution $\bm{X}^{*}$
to Problem $\eqref{eq:opt_problem}$. In the next two sections, we
 show how to approach the optimal cache placement strategy in the
centralized and distributed modes of Fog-RANs, respectively. 

\section{Submodular Optimization based Centralized Cache Placement Algorithm\label{sec:Submodular-Optimization}}

As a powerful tool for solving combinatorial optimization problems,
the submodular optimization is applied when Fog-RANs operate in the
centralized mode with the aid of a central controller. In this section,
Problem $\eqref{eq:opt_problem}$ is first reformulated into a monotone
submodular optimization problem subject to a matroid constraint. A
centralized low-complexity greedy algorithm is then proposed to obtain
a suboptimal cache placement strategy with guaranteed performance.
The basic concepts about matroid and submodular function can be found
in \cite{2003_AlexanderSchrijver_combopt}.

\subsection{Matroid Constrained Submodular Optimization}

We first define the ground set for cache placement as 
\begin{equation}
\mathcal{S}=\left\{ f_{1}^{(1)},\cdots,f_{N}^{(1)},\cdots,f_{1}^{(M)},\cdots,f_{N}^{(M)}\right\} ,
\end{equation}
where $f_{n}^{\left(m\right)}$ denotes the event that file $f_{n}$
is placed in the cache of BS $a_{m}$. The ground set $\mathcal{S}$
contains all possible caching strategies which can be applied in the
system. In particular, we use
\begin{equation}
\mathcal{S}_{m}=\left\{ f_{1}^{(m)},f_{2}^{(m)},\cdots,f_{N}^{(m)}\right\} (\forall m=1,2,\ldots,M)
\end{equation}
to denote the set of all files that might be placed in the cache of
BS $a_{m}$. Thus, the ground set $\mathcal{S}$ can be partitioned
into $M$ disjoint sets, i.e., $\mathcal{S}=\bigcup_{m=1}^{M}\mathcal{S}_{m}$,
$\mathcal{S}_{m}\bigcap\mathcal{S}_{m^{'}}=\emptyset$ for any $m\neq m^{'}$. 

Given the finite ground set $\mathcal{S}$, we continue to define
a partition matroid $\mathcal{M}=\left(\mathcal{S};\mathcal{I}\right)$,
where $\mathcal{I}\subseteq2^{\mathcal{S}}$ is a collection of independent
sets defined as: 
\begin{equation}
\mathcal{I}=\left\{ \mathcal{X}\subseteq\mathcal{S}:\left|\mathcal{X}\bigcap\mathcal{S}_{m}\right|\leq Q_{m},\forall m=1,2,\ldots,M\right\} ,\label{eq:partition_matroid}
\end{equation}
which accounts for the constraint on the cache capacity $Q_{m}$ at
each BS $a_{m}$ (c.f. ($\ref{eq:opt_problem}$.a)). The set of files
placed in the cache of BS $a_{m}$ can be denoted by $\mathcal{X}_{m}=\mathcal{X}\bigcap\mathcal{S}_{m}$.

Then, we show that the average delay is a monotone supermodular set
function over the ground set $\mathcal{S}$. Note that every set has
an equivalent boolean presentation. For any $\mathcal{X}\subseteq\mathcal{S}$,
the incidence vector of $\mathcal{X}$ is denoted by the vector $\bm{\mu}\in\left\{ 0,1\right\} ^{\mathcal{S}}$
whose $i$-th element is defined as 
\begin{equation}
\mu_{i}\doteq x_{nm},\quad i=\left(m-1\right)N+n,\label{eq:mu_x_map}
\end{equation}
where $\doteq$ represents the mapping between $x_{nm}$ and $\mu_{i}$.
In the set $\mathcal{X}\subseteq\mathcal{S}$, $f_{n}^{(m)}\in\mathcal{X}$
indicates $\mu_{i}=x_{nm}=1$. Otherwise, $\mu_{i}=x_{nm}=0$. Similarly,
the boolean presentation of the subset $\mathcal{X}_{m}$ is denoted
by $\bm{\mu}_{m}$. In this context, the delay function $\bar{D}_{nk}\left(\bm{X}\right)$
is equivalent to the set function $\bar{D}_{nk}\left(\mathcal{X}\right)$
over the set $\mathcal{X}\subseteq\mathcal{S}$. The property of $\bar{D}_{nk}\left(\mathcal{X}\right)$
is summarized in the following theorem. 
\begin{thm}
\label{thm:submodular_delay}$\tilde{D}_{nk}\left(\mathcal{X}\right)=-\bar{D}_{nk}\left(\mathcal{X}\right)$
is a monotone submodular function defined over $\mathcal{X}\in\mathcal{I}$. 
\end{thm}
\begin{IEEEproof}
The proof is deferred to Appendix \ref{subsec:Submodular-Delay}. 
\end{IEEEproof}
From \cite{2003_AlexanderSchrijver_combopt}, the class of submodular
functions is closed under non-negative linear combinations. Therefore,
for $p_{nk}\geq0$ with $k=1,2,\ldots,K$ and $n=1,2,\ldots,N$, the
set function
\begin{equation}
\tilde{D}\left(\mathcal{X}\right)=\frac{1}{K}\sum_{k=1}^{K}\sum_{n=1}^{N}p_{nk}\tilde{D}_{nk}\left(\mathcal{X}\right)
\end{equation}
is also monotone submodular. 

By taking the partition matroid $\mathcal{M}=\left(\mathcal{S};\mathcal{I}\right)$
(c.f. (\ref{eq:partition_matroid})) into consideration, Problem (\ref{eq:opt_problem})
can be reformulated into a matroid constrained monotone submodular
maximization problem:
\begin{equation}
\begin{split}\text{maximize}\quad & \tilde{D}\left(\mathcal{X}\right)=\frac{1}{K}\sum_{k=1}^{K}\sum_{n=1}^{N}p_{kn}\tilde{D}\left(\mathcal{X}\right)\\
\text{subject to}\quad & \mathcal{X}\in\mathcal{I},
\end{split}
\label{eq:opt_submodular}
\end{equation}
where the constraint $\mathcal{X}\in\mathcal{I}$ (c.f. (\ref{eq:partition_matroid}))
shows that each BS $a_{m}$ can cache up to $Q_{m}$ files. 

\subsection{Centralized Algorithm Design for Cache Placement}

\begin{algorithm}[tp]
\caption{Centralized algorithm for cache placement\label{alg:greedy}}

\begin{algorithmic}[1]

\STATE Set $\mathcal{X}\leftarrow\emptyset$ and $\mathcal{Y}\leftarrow\mathcal{S}$;

\STATE Set $\mathcal{X}_{m}\leftarrow\emptyset$ and $\mathcal{Y}_{m}\leftarrow\mathcal{S}_{m}$
for $m=1,2,\cdots,M$;		

\STATE Calculate $\Delta_{\mathcal{X}}\left(s\right)$ for each element
$s\in\mathcal{S}\backslash\mathcal{X}$;

\REPEAT 

\STATE Select the element $f_{n}^{\left(m\right)}$ with the highest
marginal gain, \\\quad $f_{n}^{\left(m\right)}=\arg\max\limits _{s\in\mathcal{S}\backslash\mathcal{X},\mathcal{X}\bigcup\left\{ s\right\} \in\mathcal{I}}\Delta_{\mathcal{X}}\left(s\right)$;

\STATE Add $f_{n}^{\left(m\right)}$ to the sets $\mathcal{X}$ and
$\mathcal{X}_{m}$:\\\quad $\mathcal{X}\leftarrow\mathcal{X}\bigcup\{f_{n}^{\left(m\right)}\}$,
$\mathcal{X}_{m}\leftarrow\mathcal{X}_{m}\bigcup\{f_{n}^{\left(m\right)}\}$;

\STATE Remove $f_{n}^{\left(m\right)}$ from the sets $\mathcal{Y}$
and $\mathcal{Y}_{m}$:\\\quad $\mathcal{Y}_{m}\leftarrow\mathcal{Y}_{m}\setminus\{f_{n}^{\left(m\right)}\}$,
$\mathcal{Y}\leftarrow\mathcal{Y}\setminus\{f_{n}^{\left(m\right)}\}$;

\IF{ $\left|\mathcal{X}_{m}\right|=Q_{m}$ }

\STATE $\mathcal{Y}\leftarrow\mathcal{Y}\setminus\mathcal{Y}_{m}$;

\ENDIF			

\STATE Calculate $\Delta_{\mathcal{X}}\left(s\right)$ for each element
$s\in\mathcal{S}\backslash\mathcal{X}$;

\UNTIL{$\mathcal{Y}=\emptyset$ or $\Delta_{\mathcal{X}}\left(s\right)=0$
for all $s\in\mathcal{S}\backslash\mathcal{X}$}

\end{algorithmic}
\end{algorithm}
We adopt a greedy algorithm \cite{2003_AlexanderSchrijver_combopt}
to find a suboptimal solution to Problem (\ref{eq:opt_submodular})
in a centralized way. Define the marginal gain of adding one element
$s\in\mathcal{S}\backslash\mathcal{X}$ to the set $\mathcal{X}$
as 
\begin{equation}
\Delta_{\mathcal{X}}\left(s\right)=\tilde{D}\left(\mathcal{X}\bigcup\left\{ s\right\} \right)-\tilde{D}\left(\mathcal{X}\right).
\end{equation}
At first, $\mathcal{X}$ and $\mathcal{X}_{m}$ are initialized to
be the empty set $\emptyset$, while $\mathcal{Y}$ and $\mathcal{Y}_{m}$
are initialized as the set $\mathcal{S}$. In each step, we calculate
the marginal gain $\Delta_{\mathcal{X}}(s)$ for each element $s\in\mathcal{S}\backslash\mathcal{X}$
and select the element $f_{n}^{(m)}$ with the highest marginal gain,
i.e., 
\begin{equation}
f_{n}^{(m)}=\arg\max_{s\in\mathcal{S}\backslash\mathcal{X},\mathcal{X}\bigcup\left\{ s\right\} \in\mathcal{I}}\Delta_{\mathcal{X}}\left(s\right),
\end{equation}
where $\mathcal{X}\bigcup\left\{ s\right\} \in\mathcal{I}$ indicates
that adding the new element $f_{n}^{(m)}$ into the current set $\mathcal{X}$
does not violate the cache capacity constraint at each BS $a_{m}$.
Then, we add this element $f_{n}^{(m)}$ to the set $\mathcal{X}_{m}$
as well as the set $\mathcal{X}$, and remove it from the sets $\mathcal{Y}$
and $\mathcal{Y}_{m}$ at the same time. When the set $\mathcal{X}_{m}$
has accumulated $Q_{m}$ elements, the set $\mathcal{Y}_{m}$  be
removed from the set $\mathcal{Y}$, which means that BS $a_{m}$
has cached up to $Q_{m}$ files and has no space for any more file.
This step runs repeatedly  until no more element can be added, i.e.,
the marginal value $\Delta_{\mathcal{X}}(s)$ is zero for all $s\in\mathcal{S}\backslash\mathcal{X}$
or the set $\mathcal{Y}$ becomes empty. The above procedures are
summarized in Algorithm $\ref{alg:greedy}$. According to \cite{j._leskovec_cost-effective_2007},
the greedy algorithm can achieve the expected $1/2$-ratio of the
optimal value in general. The computation complexity of the centralized
algorithm can be estimated as $\mathcal{O}(NMK)$ in the worst case. 

\section{Belief Propagation based Distributed Cache Placement Algorithm\label{sec:Distributed-Caching}}

 When Fog-RANs operate in the distributed mode, there exists no central
controller. The BSs should carry out a distributed algorithm for cache
placement autonomously, relying on locally collected network-side
and user-related information, as well as local interactions between
BSs in the neighborhood. In this section, we propose a belief propagation
based distributed algorithm to perform cooperative caching. The basic
concept of the message passing procedure can be found in Appendix
\ref{subsec:message_passing}.

\subsection{Factor Graph Model for Cache Placement }

To apply the belief propagation based distributed algorithm, Problem
(\ref{eq:opt_problem}) is first transformed into an unconstrained
optimization problem as presented in Lemma \ref{lem:bp_optproblem}.
To this end, we define two functions of the caching strategy $\bm{X}$
as:
\begin{equation}
\eta_{nk}(\bm{X})=\exp\left(-p_{nk}\bar{D}_{nk}(\bm{X})\right),\label{eq:obj_fun}
\end{equation}
\begin{equation}
g_{m}(\bm{X})=\begin{cases}
1, & \sum_{n=1}^{N}x_{nm}\leq Q_{m},\\
0, & \text{otherwise}.
\end{cases}\label{eq:constraint_fun}
\end{equation}

\begin{lem}
\label{lem:bp_optproblem}Let $\mathcal{C}=\{(f_{n},u_{k})|p_{nk}>0,\,f_{n}\in\mathcal{F},\,u_{k}\in\mathcal{U}\}$
denote the set of all possible pairs of file $f_{n}$ and user $u_{k}$.
Problem (\ref{eq:opt_problem}) is equivalent to the following problem
\begin{equation}
\hat{\bm{X}}=\arg\max_{\bm{X}\in\{0,1\}^{NM}}\prod_{(f_{n},u_{k})\in\mathcal{C}}\eta_{nk}(\bm{X})\prod_{m=1}^{M}g_{m}(\bm{X}).\label{eq:opt_problem_v2}
\end{equation}
\end{lem}
\begin{IEEEproof}
Problem (\ref{eq:opt_problem}) is equivalent to maximizing $-\sum_{k=1}^{K}\sum_{n=1}^{N}p_{nk}\bar{D}_{nk}(\bm{X})$
subject to the constraints $\sum_{n=1}^{N}x_{nm}\leq Q_{m}$ for all
$m$. By introducing the exponential function $\eta_{nk}(\bm{X})$
given by (\ref{eq:obj_fun}) and the indicator function $g_{m}(\bm{X})$
given by (\ref{eq:constraint_fun}), the equivalent optimization problem
is converted into a product form, as presented in (\ref{eq:opt_problem_v2}). 
\end{IEEEproof}
In (\ref{eq:opt_problem_v2}), $\eta_{nk}(\bm{X})$ is used to measure
the delay performance when transmitting file $f_{n}$ to user $u_{k}$,
and $g_{m}(\bm{X})$ imposes a strict constraint on the cache capacity
of BS $a_{m}$.

\begin{figure}
\centering
\renewcommand{\figurename}{Fig.}\subfigure[Connectivity between the BSs and users]{ \includegraphics[width=0.35\textwidth]{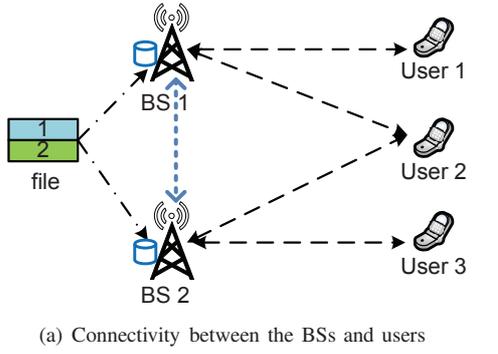}}\\\centering\subfigure[The factor graph model]{\includegraphics[width=0.4\textwidth]{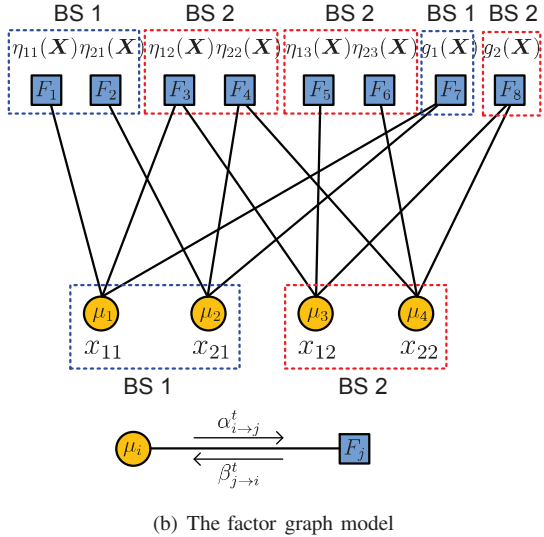}}\caption{An illustrative example: (a) a system with $2$ BSs, $3$ users, and
a library of $2$ files, (b) the factor graph model. \label{fig:factor_graph_maping}}
\end{figure}

Then, we present the factor graph model for the optimization problem
(\ref{eq:opt_problem_v2}). According to the network topology (e.g.,
Fig.$\,$\ref{fig:factor_graph_maping}(a)), we introduce a variable
node $\mu_{i}$ for each element $x_{nm}$ and a function node $F_{j}$
for each function $\eta_{nk}(\bm{X})$ or $g_{m}(\bm{X})$, as shown
in Fig.$\,$\ref{fig:factor_graph_maping}(b). The mapping rule from
$x_{nm}$ to $\mu_{i}$ is given by (\ref{eq:mu_x_map}), and the
mapping rule from $\eta_{nk}(\bm{X})$ or $g_{m}(\bm{X})$ to $F_{j}$
is expressed as
\begin{equation}
F_{j}\doteq\begin{cases}
\eta_{nk}, & j=\sum_{l=1}^{k-1}|\mathcal{F}_{l}|+\xi(n,k),\\
g_{m}, & j=\sum_{k=1}^{K}|\mathcal{\mathcal{F}}_{k}|+m,
\end{cases}
\end{equation}
where $\mathcal{\mathcal{F}}_{k}=\{f_{n}|p_{nk}>0\}$ denotes the
set of files which may be requested by user $u_{k}$, and $|\mathcal{\mathcal{F}}_{k}|$
is the number of elements in the set $\mathcal{F}_{k}$, and $\xi(n,k)$
denotes the index of file $f_{n}$  in the set $\mathcal{F}_{k}$. 

In the bipartite factor graph (e.g., Fig.$\,$\ref{fig:factor_graph_maping}(b)),
each variable node $\mu_{i}\doteq x_{nm}$ is adjacent to the function
nodes $\{F_{j}\}\doteq\{\eta_{nk}\}\bigcup\{g_{m}\}$ for all $u_{k}\in\mathcal{U}_{m}$.
Similarly, each function node $F_{j}\doteq\eta_{nk}$ is connected
to the variable nodes $\{\mu_{i}=x_{nm}\}$ for all $a_{m}\in\mathcal{A}_{k}$.
Each function node $F_{j}\doteq g_{m}$ is adjacent to the variable
nodes $\{\mu_{i}\doteq x_{nm}\}$ for all $f_{n}\in\mathcal{F}$.
Hence, there are $I=NM$ variable nodes and $J=M+\sum_{k=1}^{K}|\mathcal{F}_{k}|$
function nodes in this factor graph model.  

\subsection{Message Passing Procedure for Cache Placement \label{subsec:Practical-Message-Passing}}

Our goal is to design a message-passing procedure which allows us
to gradually approach the optimal solution to (\ref{eq:opt_problem_v2}).

\subsubsection{Message Update }

Let $m_{\mu_{i}\rightarrow F_{j}}^{t}(x)$ denote the message from
a variable node $\mu_{i}$ to a function node $F_{j}$, and $m_{F_{j}\rightarrow\mu_{i}}^{t}(x)$
denote the message from a function node $F_{j}$ to a variable node
$\mu_{i}$, respectively. The update of the messages $m_{\mu_{i}\rightarrow F_{j}}^{t}(x)$
and $m_{F_{j}\rightarrow\mu_{i}}^{t}(x)$ can be obtained by (\ref{eq:variable-function})
and (\ref{eq:function-variable}), respectively. Since all the variables
$\{x_{nm}\}$ are binary, it is sufficient to pass the scalar ratio
of the messages between each pair of nodes in practice.  We can
also express the message ratios in the logarithmic domain as
\begin{equation}
\alpha_{i\rightarrow j}^{t}=\log\left(\frac{m_{\mu_{i}\rightarrow F_{j}}^{t}(1)}{m_{\mu_{i}\rightarrow F_{j}}^{t}(0)}\right),\beta_{j\rightarrow i}^{t}=\log\left(\frac{m_{F_{j}\rightarrow\mu_{i}}^{t}(1)}{m_{F_{j}\rightarrow\mu_{i}}^{t}(0)}\right).\label{eq:message_alpha}
\end{equation}
In this way, the computation complexity and communication overhead
are greatly reduced. This is because only half of the messages are
actually calculated and passed. As shown in Fig.$\,$\ref{fig:factor_graph_maping}(b),
the message $\alpha_{i\rightarrow j}^{t}$, instead of $m_{\mu_{i}\rightarrow F_{j}}^{t}(x)$
$(x\in\{0,1\})$, is sent from the variable node $\mu_{i}$ to the
function node $F_{j}$, and the message $\beta_{j\rightarrow i}^{t}$,
instead of $m_{F_{j}\rightarrow\mu_{i}}^{t}(x)$ $(x\in\{0,1\})$,
is sent from the function node $F_{j}$ to the variable node $\mu_{i}$.
Meanwhile, the product operations in (\ref{eq:variable-function})
and (\ref{eq:function-variable}) become simple additive operations
in the logarithmic domain, as presented in the following theorem.

\begin{thm}
\label{thm:message_update}The message $\alpha_{i\rightarrow j}^{t}$
is updated as
\begin{equation}
\alpha_{i\rightarrow j}^{t+1}=\sum_{l\in\Gamma_{i}^{\mu}\backslash\{j\}}\beta_{l\rightarrow i}^{t}.\label{eq:alpha_variable_function_ratio}
\end{equation}
When  $F_{j}\doteq\eta_{nk}$, the message $\beta_{j\rightarrow i}^{t+1}$
is given by
\begin{equation}
\beta_{j\rightarrow i}^{t+1}=p_{nk}\left(\bar{D}_{nk}(\bm{X}_{i,0}^{t})-\bar{D}_{nk}(\bm{X}_{i,1}^{t})\right),\label{eq:beta_function_variable_ratio}
\end{equation}
where the caching vectors $\bm{X}_{i,0}^{t}$ and \textup{$\bm{X}_{i,1}^{t}$}
can be obtained by \textup{\emph{assigning their elements as}} 
\[
x_{nm}\doteq\mu_{l}=\begin{cases}
1, & l\in E_{i}^{t}=\{i_{1}\in\Gamma_{j}^{F}\backslash\{i\}|\alpha_{i_{1}\rightarrow j}^{t}>0\},\\
0, & \text{otherwise},
\end{cases}
\]
and
\[
x_{nm}\doteq\mu_{l}=\begin{cases}
1, & l\in E_{i}^{t}\bigcup\{i\},\\
0, & \text{otherwise},
\end{cases}
\]
respectively. When  $F_{j}\doteq g_{m}$, the message $\beta_{j\rightarrow i}^{t}$
is updated as 
\begin{equation}
\beta_{j\rightarrow i}^{t+1}=\min\left\{ 0,-\alpha_{l\rightarrow j}^{(Q_{m})}(t)\right\} ,\label{eq:xi_function_variable_ratio}
\end{equation}
where $\alpha_{l\rightarrow j}^{(Q_{m})}(t)$ is the $Q_{m}$-th message
among the messages $\{\alpha_{l\rightarrow j}^{t}\}$ $(l\in\Gamma_{j}^{F}\backslash\{i\})$
sorted in the descending order.
\end{thm}
\begin{IEEEproof}
The proof is deferred to Appendix \ref{subsec:Proof-of-Theorem-message-update}.\emph{}
\end{IEEEproof}
In practice, the messages $\alpha_{i\rightarrow j}^{t}$ and $\beta_{j\rightarrow i}^{t}$
reflect the beliefs on the value of $\mu_{i}$ and should be updated
according to (\ref{eq:alpha_variable_function_ratio}) and (\ref{eq:beta_function_variable_ratio})
(or (\ref{eq:xi_function_variable_ratio})), respectively, in each
iteration.

\subsubsection{Belief Update }

In the $t$-th iteration, the belief on $\mu_{i}=x$ is expressed
as
\begin{equation}
b_{i}^{t+1}(x)=\prod_{j\in\Gamma_{i}^{\mu}}m_{F_{j}\rightarrow\mu_{i}}^{t}(x),\label{eq:belief}
\end{equation}
which is the product of all the messages incident to $\mu_{i}$. Hence,
the belief ratio in the logarithmic domain can be obtained as 
\begin{equation}
\tilde{b}_{i}^{t}=\log\left(\frac{b_{i}^{t}(1)}{b_{i}^{t}(0)}\right)=\sum_{j\in\Gamma_{i}^{\mu}}\beta_{j\rightarrow i}^{t},\label{eq:belief-ratio}
\end{equation}
where $\beta_{j\rightarrow i}^{t}$ is given by (\ref{eq:xi_function_variable_ratio})
for $F_{l}\doteq g_{m}$, and by (\ref{eq:beta_function_variable_ratio})
for $F_{j}\doteq\eta_{nk}$ ($j\in\Gamma_{i}^{\mu}\backslash\{l\}$),
respectively. As a result, the estimation of $\mu_{i}$ can be expressed
as 
\begin{equation}
\hat{\mu}_{i}^{t}=\begin{cases}
1, & \text{if }\tilde{b}_{i}^{t}>0,\\
0, & \text{if }\tilde{b}_{i}^{t}<0.
\end{cases}\label{eq:decision-v2}
\end{equation}
In each iteration, each variable node $\mu_{i}$ updates its belief
on its associated variable $x_{nm}$ according to (\ref{eq:belief-ratio})
and makes an estimate of $x_{nm}$ according to (\ref{eq:decision-v2})
until it converges.
\begin{algorithm}[t]
\caption{Distributed algorithm for cache placement\label{alg:bp_algorithm}}

\begin{algorithmic}[1]

\STATE Map $\eta_{nk}$, $g_{m}$ to $F_{j}$ and $x_{nm}$ to $\mu_{i}$
for $\forall n,k,m$, 

\STATE Set $t=0$ and $\alpha_{i\rightarrow j}^{t}=\beta_{j\rightarrow i}^{t}=0$,
$\forall i,j$,

\STATE Set $t_{max}$ as a sufficiently large constant.

\WHILE{ Not convergent and $t\leq t_{max}$}

\FOR{$m=1:M$}

\FOR{$n=1:N$}

\STATE Calculate the message $\alpha_{i\rightarrow j}^{t}$ by (\ref{eq:alpha_variable_function_ratio});

\FOR{$k\in\widetilde{\mathcal{U}}_{m}$}

\STATE Calculate the message $\beta_{j\rightarrow i}^{t}$ for $F_{j}\doteq\eta_{nk}$
by (\ref{eq:beta_function_variable_ratio});

\ENDFOR 

\ENDFOR

\STATE Calculate the message $\beta_{j\rightarrow i}^{t}$ for $F_{j}\doteq g_{m}$
by (\ref{eq:xi_function_variable_ratio});

\STATE Calculate the belief $\tilde{b}_{i}^{t}$ by (\ref{eq:belief-ratio});

\STATE Estimate each variable $\hat{\mu}_{i}$ by (\ref{eq:decision-v2});

\ENDFOR

\STATE Check the convergence, and set $t=t+1$;

\ENDWHILE

\STATE Obtain the optimal estimate $\hat{\bm{X}}$ to the solution
of (\ref{eq:opt_problem_v2}). 

\end{algorithmic}
\end{algorithm}

\subsection{Distributed Cache Placement Algorithm }

When we map the message passing procedure derived on the factor graph
(e.g., Fig.$\,$\ref{fig:factor_graph_maping}(b)) back to the original
network graph (e.g., Fig.$\,$\ref{fig:factor_graph_maping}(a)),
we notice that all the messages are updated at the BSs and some of
them  be exchanged between neighboring BSs.

\subsubsection{Scenario I}

When user $u_{k}$ is connected to one single BS $a_{m}$, as shown
in Fig.$\,$\ref{fig:factor_graph_maping}(b), the update of messages
$\alpha_{i\rightarrow j}^{t}$ and $\beta_{j\rightarrow i}^{t}$ is
performed at this BS for the variable node $\mu_{i}\doteq x_{nm}$,
the function nodes $F_{j}\doteq\eta_{nk}$, and $F_{j}\doteq g_{m}$.
In this case, each BS $a_{m}$ performs the message calculation and
belief update for all the users just served by itself, i.e., $u_{k}\in\mathcal{U}_{m}$
and $|\mathcal{A}_{k}|=1$. 

\subsubsection{Scenario II}

When user $u_{k}$ is in the coverage of multiple BSs $\mathcal{A}_{k}$,
the update of messages $\alpha_{i\rightarrow j}^{t}$ and $\beta_{j\rightarrow i}^{t}$
associated with the function node $F_{j}\doteq\eta_{nk}$ is performed
at one BS $a_{m}$ and  be exchanged between the serving BSs of this
user $\mathcal{A}_{k}$ over control links, as shown in Fig.$\,\ref{fig:system_model}$. 

Notice that message exchanges just take place in Scenario II, and
the communication overhead induced depends on the number of common
users covered by multiple BSs. From the above discussion, we summarize
the message passing based distributed algorithm for cache placement
in Algorithm \ref{alg:bp_algorithm}. In this algorithm, the message
update for each user should be performed just once by one single BS
in each iteration.  To avoid confusion, $\tilde{\mathcal{U}}_{m}$
is used to denote the set of users whose messages are processed by
BS $a_{m}$ in Algorithm \ref{alg:bp_algorithm}. 
\begin{figure}[tp]
\centering
\renewcommand{\figurename}{Fig.}\subfigure[The average download delay]{\label{fig:delay_Q_gamma065}
\includegraphics[width=0.5\textwidth]{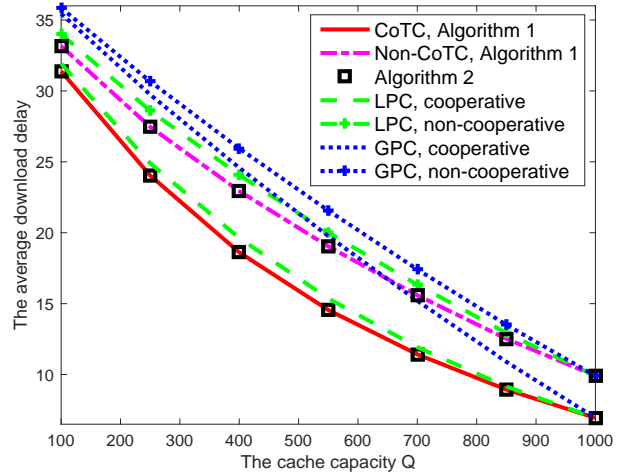}}\\\centering
\renewcommand{\figurename}{Fig.}\subfigure[The average hit probability]{\label{fig:hit_Q_gamma065}
\includegraphics[width=0.5\textwidth]{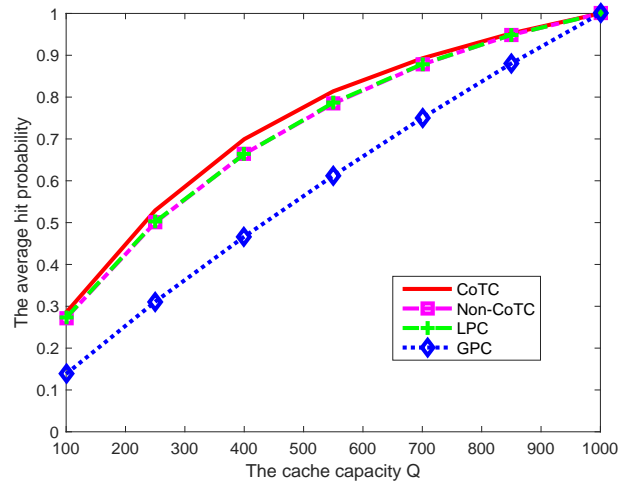}}\caption{The average delay and hit probability of the proposed caching strategies
when $\gamma_{k}=0.65$ and $N=1000$. \label{fig:gamma065}}
\end{figure}

\section{Simulation Results\label{sec:Simulation-Results}}

In this section, we present simulation results to demonstrate the
performance of the proposed cache placement algorithms, i.e., Algorithm
\ref{alg:greedy} and Algorithm \ref{alg:bp_algorithm}. We consider
a Fog-RAN with $M$ BSs and $K$ mobile users. Each BS serves the
users in a circular cell with a radius of $150$m, and the distance
between neighboring BSs is $200$m. $K$ users are uniformly and independently
distributed in the area covered by the $M$ cells. File requests of
each user $u_{k}$ follow the Zipf distribution with parameter $\gamma_{k}$.
The users in the cell interior are served by just one single BS, while
the users in the overlapping area of cells are covered by multiple
BSs and thus cooperative transmission may be enabled. The connectivity
between the BSs and users is thus established.

Suppose that the system bandwidth is $5$MHz, and the length of each
time slot is $20$ms. The file size is equal to $100$Mbits. The path-loss
exponent is set as $3.5$. The small-scale channel gain $|h_{km}|^{2}$
follows independently standard exponential distribution in each time
slot. Assume that no inter-cell interference is induced by adopting
appropriate scheduling policies, and the transmit power is set to
make sure that the average received SNR at the cell edge is equal
to $0$dB. Unless otherwise stated, we set $K=100$, $M=10$, and
$D_{nk}=40$s. Suppose that each user $u_{k}$ requests file $f_{n}$
with probability $p_{nk}=\frac{(\phi(n))^{-\gamma_{k}}}{\sum_{n=1}^{N}n^{-\gamma_{k}}}$,
where $\{\phi(n)\}_{n=1}^{N}$ is a random permutation of $[1,\cdots,N]$,
i.e., we assume different users have different request distributions. 

In the considered system, we  compare two transmission aware caching
strategies and two baseline popular caching strategies: 1) Non-cooperative
transmission aware caching (Non-CoTC) strategy, which is designed
based on prior knowledge that each individual user has the file preference
$p_{nk}$ and is served by one serving BS using non-cooperative transmission
given by (\ref{eq:fixedService}); 2) Cooperative transmission aware
caching (CoTC) strategy, which is designed based on prior knowledge
that each individual user has the file preference $p_{nk}$ and is
served by one BS using non-cooperative transmission given by (\ref{eq:fixedService}),
or by multiple BSs using cooperative beamforming given by (\ref{eq:Beamforming}),
depending on the connectivity between the user and the BSs; 3) Globally
popular caching (GPC) strategy, which caches the most $Q_{m}$ popular
files at each BS $a_{m}$ based on the network-wide file popularity
$\{\tilde{p}_{n}\}$. Here, the file popularity is evaluated as $\tilde{p}_{n}=\frac{1}{K}\sum_{k=1}^{K}p_{nk}$,
i.e., the average value of the file preferences of users in the network;
4) Locally popular caching (LPC) strategy, which caches the most $Q_{m}$
popular files at each BS $a_{m}$ based on the local file popularity
$\tilde{p}_{n}^{(m)}=\frac{1}{|\mathcal{U}_{m}|}\sum_{u_{k}\in\mathcal{U}_{m}}p_{nk}$,
i.e., the average value of the file preferences of users served by
the BS $a_{m}$.  The proposed transmission aware caching strategies
can be performed in either a centralized or a distributed way. There
is no difference between centralized and distributed ways of performing
the Popular caching strategy.

\subsection{Performance Evaluation}

\begin{figure}[tp]
\centering
\renewcommand{\figurename}{Fig.}\subfigure[The average download delay]{\label{fig:delay_Q_gamma025}
\includegraphics[width=0.5\textwidth]{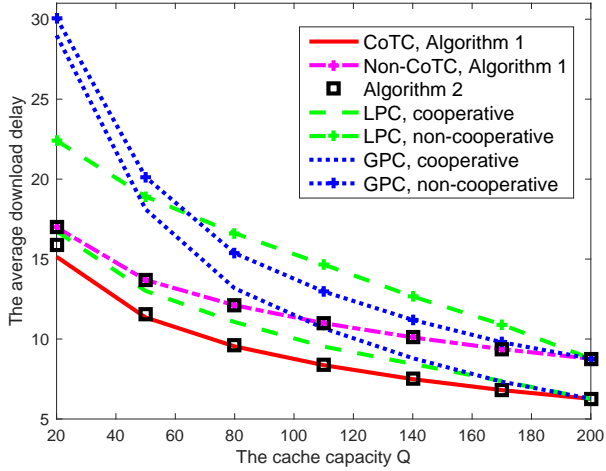}}\\\centering
\renewcommand{\figurename}{Fig.}\subfigure[The average hit probability]{\label{fig:hit_Q_gamma025}
\includegraphics[width=0.5\textwidth]{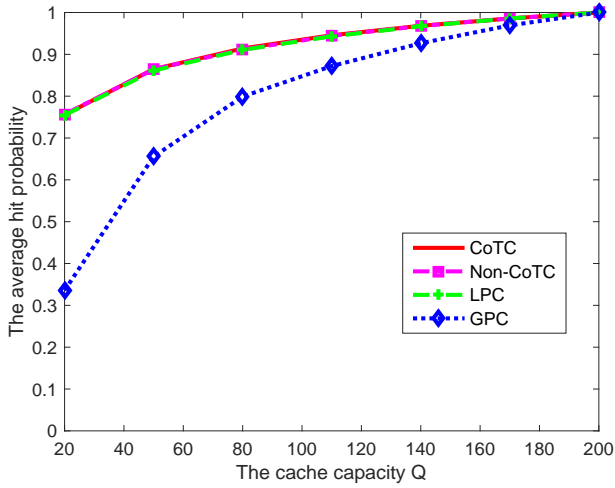}}\caption{The average delay and hit probability performances of the proposed
caching strategies when $\gamma_{k}=0.2+4.8\frac{k}{K}$ and $N=200$.\label{fig:gamma025}}
\end{figure}

We demonstrate the performances of our considered four caching strategies
in two scenarios when $\gamma_{k}=0.65$, $N=1000$ and $\gamma_{k}=0.2+4.8\frac{k}{K}$,
$N=200$ in Fig.$\,$\ref{fig:gamma065} and Fig.$\,$\ref{fig:gamma025},
respectively. In each scenario, we plot the average download delay
and hit probability curves of these caching strategies in sub-figures
(a) and (b), respectively, for different cache capacities $Q_{m}=Q$.
When our proposed Non-CoTC or CoTC strategy is applied, the users'
average download delay $\bar{D}(\bm{X})$ is computed by substituting
the solution $\bm{X}$ that is achieved either by Algorithm 1 or by
Algorithm 2. When the GPC or LPC strategy is applied, the average
delay $\bar{D}(\bm{X})$ is obtained by substituting the GPC or LPC
solution $\bm{X}$. As shown in Fig.$\,$\ref{fig:gamma065} and Fig.$\,$\ref{fig:gamma025},
the average download delay monotonically decreases with the increase
of the cache capacity $Q$ for any given caching strategy. This is
due to the fact that with the increase of storage capacity, more files
are cached in each BS and more users can download files from local
BSs instead of the content server. Due to the same reason, the users'
average hit probability monotonically increases with the cache capacity. 

As shown in Fig.$\,$\ref{fig:delay_Q_gamma065} and Fig.$\,$\ref{fig:delay_Q_gamma025},
the two transmission aware caching strategies, i.e., Non-CoTC and
CoTC, achieve smaller average download delays than the two popular
caching strategies, i.e., LPC and GPC, for any cache capacity $Q$
less than $N$. Meanwhile, the average hit probabilities of the CoTC
and Non-CoTC strategies are higher or equal to that of the LPC strategy,
and much higher than the GPC strategies when $Q<N$, as shown in Fig.$\,$\ref{fig:hit_Q_gamma065}
and Fig.$\,$\ref{fig:hit_Q_gamma025}. This is because the transmission
aware caching strategies cache files at the BSs based on the accurate
file preferences of individual users and the prior information on
content delivery techniques that will be applied by the BSs. While
the LPC or GPC strategy performs caching based on the file preference
statistics of the users in each cell or in the network, which could
not reflect the file preferences of individual users. 

The delay performance of the caching strategies not only depends on
the users' hit performance, but also on the transmission schemes the
BSs will adopt to deliver the requested files. It is observed from
Fig.$\,$\ref{fig:delay_Q_gamma065} and Fig.$\,$\ref{fig:delay_Q_gamma025}
that the CoTC strategy performs much better than the Non-CoTC strategy
in terms of the average delay and hit probability. The delay performance
gap between the two transmission aware caching strategies becomes
larger as the cache capacity increases, since more files can be cached
to facilitate cooperative transmission for cell-edge users. In other
words, the CoTC strategy can exploit both caching gain and cooperative
gain to reduce the average delay. Hence, the design of caching strategies
should not only target at improving the users' average hit probability,
but also bringing more cooperative transmission opportunities. Similarly,
the delay performance is significantly improved when cooperative transmission
is applied instead of non-cooperative transmission for any caching
strategy. 

At the same time, the users' skewness on content popularity has a
great impact on the performances of the considered caching strategies.
When $\gamma_{k}=0.65$, each user is interested in a large number
of files while only a very small number of files can be cached locally
at the serving BSs of each user when $Q$ is less than $N$. From
Fig.$\,$\ref{fig:delay_Q_gamma065}, the delay gap between the CoTC
(or Non-CoTC) strategy and the LPC strategy is not very large. And
the GPC strategy which caches the same files in each BS achieves the
worst delay and hit performances. When $\gamma_{k}=0.2+4.8\frac{k}{K}$,
the skewness on content popularity is quite different among users.
This means some users have interests on many files while some users
just have preferences on very few files. In contrast to the case with
$\gamma_{k}=0.65$, a higher proportion of the files that the users
may request can be cached at the BSs. Therefore, the delay and hit
performances of the considered caching strategies are all improved.
And the delay gap between the CoTC (or Non-CoTC) strategy and the
LPC strategy becomes very significant especially when the cache capacity
$Q$ is very small. It is also interesting to see that the delay performance
of the LPC strategy gets affected by content delivery schemes applied
by the BSs. As shown in Fig.$\,$\ref{fig:delay_Q_gamma025}, the
LPC strategy always achieves a smaller average delay than the GPC
strategy if cooperative transmission is adopted. However, it performs
worse in the larger $Q$ region ($Q>65$) when non-cooperative transmission
is applied. This happens when some users are served by their serving
BSs which have not cached their requested files, since the LPC strategy
caches files based on the file preferences of co-located users and
pushes quite different contents in each BS. 

From Fig.$\,$\ref{fig:gamma065} and Fig.$\,$\ref{fig:gamma025},
the proposed belief propagation based distributed algorithm can achieve
a nearly identical delay performance as compared to the centralized
greedy algorithm which provides a guaranteed performance \cite{j._leskovec_cost-effective_2007},
i.e., $1/2$-approximation in the general case and $(1-1/e)$-approximation
in some special cases. It has a slightly larger delay performance
in the small-capacity region (e.g., $Q$ is around $20$), and achieves
almost the same performance as the centralized algorithm in other
scenarios.

\subsection{Approximation of File Preferences }

\begin{figure}[tp]
\centering
\renewcommand{\figurename}{Fig.} \includegraphics[width=0.5\textwidth]{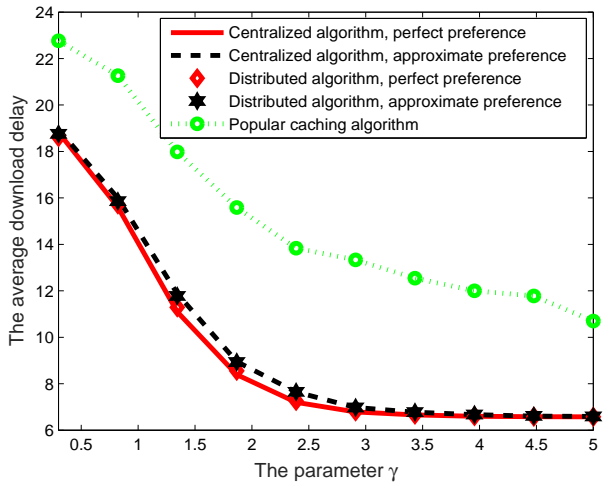}\caption{The average download delay vs. the parameter $\gamma$. \label{fig:delay_gamma}}
\end{figure}
In practice, it is very challenging to accurately estimate the file
preference of each individual user due to the lack of sufficient samples.
Instead, each BS may estimate an approximate file preference for all
the users in its coverage, i.e., to estimate the average preference.
In this part, we  discuss the impact of the users' file request preference
statistics, either perfectly or approximately known. In Fig.$\,$\ref{fig:delay_gamma},
we show how the average download delay changes with the content popularity
skewness. In this experiment, all the users are supposed to have the
same preference parameter $\gamma_{k}=\gamma$. The cache capacity
is set as $Q=50$ and the total number of files is $N=100$. The approximate
preference for file $f_{n}$ is given by $\tilde{p}_{nk}=\frac{1}{|\mathcal{U}_{m}|}\sum_{u_{k}\in\mathcal{U}_{m}}p_{nk}$
($\forall u_{k}\in\mathcal{U}_{m}$), i.e., only the statistical average
of all the users in the coverage of each BS $a_{m}$ is known, while
a perfect knowledge $p_{nk}$ includes preference for each individual
user. It is observed that the average delay is significantly reduced
when the parameter $\gamma$ is increased within $0.3\leq\gamma\leq3$.
In this range, the users have preferences on fewer and fewer files
with the increase of the parameter $\gamma$. This means that more
and more requested files are cached at the BSs, and can be transmitted
to the users directly. As a result, the average download delay is
greatly reduced when $\gamma$ is increased within $0.3\leq\gamma\leq3$.
When $\gamma>3$, almost all the requested files have been cached
and the average download delay is nearly equal to the average transmission
time from the BSs to the users. In this case, the change of the average
delay is not obvious. In Fig.$\,$\ref{fig:delay_gamma}, we also
plot the average delay performance when approximate file preferences
instead of accurate file preferences are applied. It can be seen that
the delay gap is very small. 

\begin{figure}[tp]
\centering
\renewcommand{\figurename}{Fig.} \includegraphics[width=0.5\textwidth]{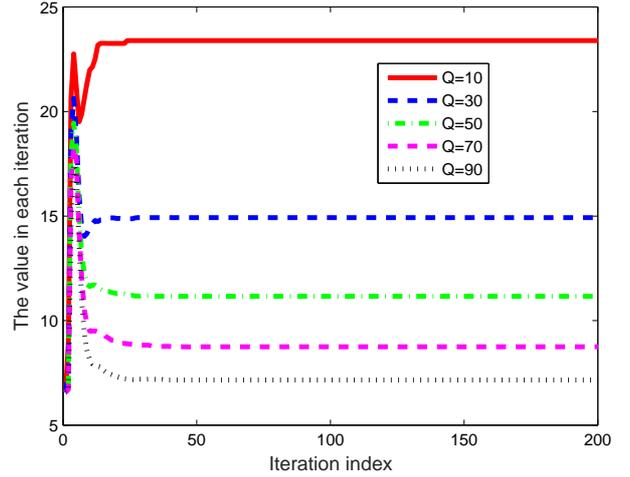}\caption{The iterative procedure of the proposed distributed algorithm. \label{fig:BP_convergence}}
\end{figure}
In Fig.$\,$\ref{fig:BP_convergence}, we plot the iterative procedure
of the belief propagation based distributed algorithm for different
storage capacities $Q_{m}=Q$ and $N=100$. In this experiment, the
 CoTC strategy is performed in a distributed way. It is observed that
the average delay starts from an initial value, fluctuates up to dozens
of iterations and gradually converges to a suboptimal solution.

\subsection{Algorithm Complexity}

\begin{figure}[tp]
\centering
\renewcommand{\figurename}{Fig.} \includegraphics[width=0.5\textwidth]{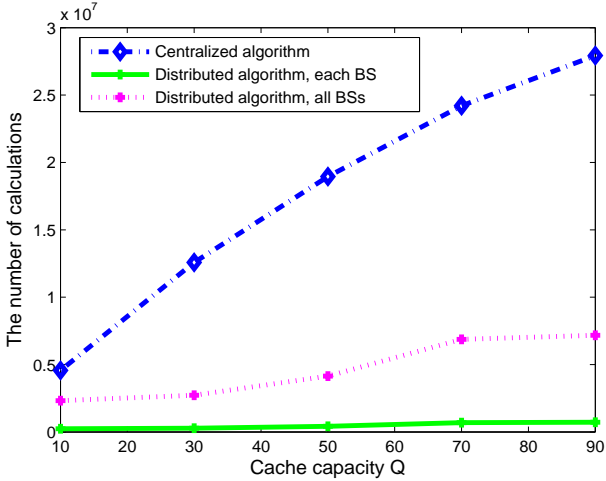}\caption{The number of calculations vs. cache capacity $Q$. \label{fig:complexity}}
\end{figure}
We now discuss the computation complexity of our proposed centralized
and distributed algorithms when performing the CoTC strategy. Here,
we measure the computation complexity by the number of calculations
required in the algorithms. In Fig.$\,$\ref{fig:complexity}, we
plot the computation complexity of the proposed algorithms versus
the cache capacity $Q$. In this experiment, the number of BSs and
the number of users are set as $M=10$ and $K=100$, and the total
number of files is set to be $100$. It can be seen that the computation
complexity of the centralized algorithm rapidly increases with the
increase of the cache capacity $Q$, while the computation complexity
of the distributed algorithm increases very slowly with the cache
capacity $Q$. This indicates that the cache capacity has a greater
impact on the computation complexity of the centralized algorithm
rather than the distributed algorithm, since more elements are added
greedily and more iterations are processed in the centralized algorithm
when the cache capacity $Q$ is increased. When applying the distributed
algorithm, the cache capacity is a parameter which only adjusts the
value of the messages during iterations. It does not change the factor
graph model, and hence may not cause a significant impact on its computation
complexity. 

\section{Conclusions\label{sec:Conclusions}}

In this work, we studied the cache placement problem in Fog-RANs,
by taking into account different file preferences and diverse transmission
opportunities for each user. We developed transmission aware cache
placement strategies in both centralized and distributed operation
modes of Fog-RANs. In the centralized mode, a low-complexity centralized
greedy algorithm was proposed to achieve a suboptimal solution within
a constant factor to the optimum using submodular optimization techniques.
In the distributed mode, a low-complexity belief propagation based
distributed algorithm was proposed to place files at the BSs based
on locally collected information. Each BS  run computations and exchange
very few messages with its neighboring BSs iteratively until convergence.
By simulations, we showed that both of the proposed algorithms can
not only improve the users' cache hit probability but also provide
more flexible cooperative transmission opportunities for the users.
As a result, our proposed centralized and distributed cache placement
algorithms can significantly improve the file delivery performance
by providing cooperative transmission opportunities for mobile users
to the maximum extent.  It was also shown that the distributed cache
placement  algorithm can achieve an average delay performance comparable
to the centralized cache placement algorithm while spending much less
calculations in each individual BS.

\appendix
\renewcommand{\appendixname}{Appendix~\Alph{section}}

\subsection{Proof of Theorem \ref{thm:avdelay} \label{subsec:Theorem1_delay}}

In the scenario when file $f_{n}$ has been cached in one or multiple
serving BSs of user $u_{k}$, i.e., $\sum_{a_{m}\in\mathcal{A}_{k}}x_{nm}\neq0$,
the associated BSs can transmit to user $u_{k}$ with rate $R_{nk}(\bm{X},i)$
(c.f. (\ref{eq:fixedService})-(\ref{eq:Beamforming})) by applying
some specific transmission scheme. Since the channel coefficients
$h_{km}(i)$ are i.i.d. across the time slots $\{i\}$, the file delivery
rates $R_{nk}(\bm{X},i)$ are i.i.d. random variables. Hence, the
stopping time of completing the transmission of file $f_{n}$, $T_{nk}^{*}(\bm{X})$
given by (\ref{eq:StoppingTime}), is also a random variable. Based
on the definition of channel capacity, we have $R_{nk}(\bm{X},i)\geq0$
for $i=1,2,\cdots,T_{nk}^{*}(\bm{X})$. According to Wald's Equation
in martingale theory \cite{david_williams_probability_1991}, we have
\begin{equation}
\begin{array}[t]{cl}
 & \mathbb{E}_{\bm{h}}\left\{ \sum\nolimits _{i=1}^{T_{nk}^{*}(\bm{X})}R_{nk}(\bm{X},i)\right\} \\
= & \mathbb{E}_{\bm{h}}\left\{ T_{nk}^{*}(\bm{X})\right\} \cdot\mathbb{E}_{\bm{h}}\left\{ R_{nk}(\bm{X})\right\} =\frac{|f_{n}|}{\Delta t}.
\end{array}
\end{equation}
Therefore, the average download delay is expressed as 
\begin{equation}
\bar{D}_{nk}(\bm{X})=\mathbb{E}_{\bm{h}}\left\{ T_{nk}^{*}(\bm{X})\cdot\Delta t\right\} =\frac{|f_{n}|}{\mathbb{E}_{\bm{h}}\left\{ R_{nk}(\bm{X})\right\} },
\end{equation}
when file $f_{n}$ is cached in the associated BSs with $\sum_{a_{m}\in\mathcal{A}_{k}}x_{nm}\neq0$.
When $\sum_{a_{m}\in\mathcal{A}_{k}}x_{nm}=0$, file $f_{n}$ has
not been cached in any serving BS of user $u_{k}$. The BSs $\mathcal{A}_{k}^{'}$
 download this file from the content server by the backhaul link and
then transmit to user $u_{k}$ over the wireless channel. Accordingly,
the average delay can be estimated by $\bar{D}_{nk}(\bm{X})=D_{nk}+\frac{|f_{n}|}{\mathbb{E}_{\bm{h}}\left\{ R_{nk}(\bm{X}_{k})\right\} }$,
where $D_{nk}$ is the extra delay of file delivery from the content
server to the serving BSs $\mathcal{A}_{k}^{'}$, and $R_{nk}(\bm{X}_{k})$
is the data rate at which the BSs $\mathcal{A}_{k}^{'}$ transmit
file $f_{n}$ to user $u_{k}$ over wireless channel. Here, $\bm{X}_{k}$
is an equivalent caching strategy indicating that file $f_{n}$ can
be downloaded from the BSs $\mathcal{A}_{k}^{'}$ by user $u_{k}$.
 Thus, the average delay $\bar{D}_{nk}(\bm{X})$ is established in
$\eqref{eq:av_Delay}$. 

\subsection{\label{subsec:Submodular-Delay}Proof of Theorem \ref{thm:submodular_delay}}

From Theorem \ref{thm:avdelay}, the average delay of downloading
file $f_{n}$ for user $u_{k}$ presented in (\ref{eq:av_Delay})
can also be expressed as 
\begin{equation}
\bar{D}_{nk}\left(\mathcal{X}\right)=\begin{cases}
\frac{\left|f_{n}\right|}{\bar{R}_{nk}(\mathcal{X})}, & \sum_{m=1}^{M}x_{nm}\neq0,\\
D_{nk}+\frac{\left|f_{n}\right|}{\bar{R}_{nk}(\mathcal{X}_{k})}, & \textrm{otherwise},
\end{cases}
\end{equation}
where $\bar{R}_{nk}(\mathcal{X})=\mathbb{E}\left\{ B\log(1+Y_{nk}(\mathcal{X}))\right\} $
with $Y_{nk}(\mathcal{X})=\sum_{m=1}^{M}|h_{km}|^{2}x_{nm}\mathrm{SINR}_{m}$
representing the received SINR. We will show that the average delay
$\tilde{D}_{nk}\left(\mathcal{X}\right)=-\bar{D}_{nk}\left(\mathcal{X}\right)$
is a monotone submodular function.

Let $\mathcal{X}\subseteq\mathcal{X}'\in\mathcal{I}$, and $s\in\mathcal{S}\setminus\mathcal{X}'$.
The incidence vectors for $\mathcal{X}$ and $\mathcal{X}^{'}$ are
denoted by $\bm{X}=[x_{nm}]$  and $\bm{X}^{'}=[x_{nm}^{'}]$, respectively.
If $s\neq f_{n}^{\left(m\right)}$ for any $m\in\mathcal{\mathcal{A}}_{k}$,
we have $\tilde{D}_{nk}\left(\mathcal{X}\cup\left\{ s\right\} \right)-\tilde{D}_{nk}\left(\mathcal{X}\right)=\tilde{D}_{nk}\left(\mathcal{X}'\cup\left\{ s\right\} \right)-\tilde{D}_{nk}\left(\mathcal{X}'\right)=0$.
We then consider the case when $s=f_{n}^{\left(m^{*}\right)}$ for
any $m^{*}\in\mathcal{\mathcal{A}}_{k}$. 

\textbf{Case I: } $\mathcal{X}=\mathcal{X}^{'}\in\mathcal{I}$ and
$\sum_{m\in\mathcal{A}_{k}}x_{nm}=\sum_{m\in\mathcal{A}_{k}}x_{nm}^{'}$

In this case, $s=\emptyset\in\mathcal{X}'\setminus\mathcal{X}$ and
$\bar{D}_{nk}\left(\mathcal{X}\right)=\bar{D}_{nk}\left(\mathcal{X}'\right)$.
Hence, we have $\tilde{D}_{nk}\left(\mathcal{X}\cup\left\{ s\right\} \right)-\tilde{D}_{nk}\left(\mathcal{X}\right)=\tilde{D}_{nk}\left(\mathcal{X}'\cup\left\{ s\right\} \right)-\tilde{D}_{nk}\left(\mathcal{X}'\right)=0$.

\textbf{Case II: } $\mathcal{X}\subseteq\mathcal{X}'\in\mathcal{I}$
and $0<\sum_{m\in\mathcal{A}_{k}}x_{nm}<\sum_{m\in\mathcal{A}_{k}}x_{nm}^{'}$

According to the definition of $\bar{R}_{nk}(\mathcal{X})$, we have
$\bar{R}_{nk}(\mathcal{X}\cup\{s\})=\mathbb{E}\{B\log(1+Y_{nk}(\mathcal{X})+|h_{km^{*}}|^{2}\mathrm{SINR}_{m^{*}}))\}$.
Hence, $\bar{R}_{nk}(\mathcal{X})<\bar{R}_{nk}(\mathcal{X}^{'})$
and $\bar{R}_{nk}(\mathcal{X}\cup\{s\})<\bar{R}_{nk}(\mathcal{X}^{'}\cup\{s\})$
naturally hold due to $\sum_{m=1}^{M}x_{nm}<\sum_{m=1}^{M}x_{nm}^{'}$
and $Y_{nk}(\mathcal{X})<Y_{nk}(\mathcal{X}^{'})$. The gap between
$\tilde{D}_{nk}\left(\mathcal{X}\cup\left\{ s\right\} \right)$ and
$\tilde{D}_{nk}\left(\mathcal{X}\right)$ satisfies
\[
\begin{split} & \tilde{D}_{nk}\left(\mathcal{X}\cup\left\{ s\right\} \right)-\tilde{D}_{nk}\left(\mathcal{X}\right)\\
= & \frac{\left|f_{n}\right|}{\bar{R}_{nk}(\mathcal{X})\bar{R}_{nk}(\mathcal{X}\cup\{s\})}\mathbb{E}\left\{ B\log\left(1+\frac{|h_{km^{*}}|^{2}\mathrm{SINR}_{m^{*}}}{1+Y_{nk}(\mathcal{X}^{'})}\right)\right\} \\
\underset{>}{(a)} & \frac{\left|f_{n}\right|}{\bar{R}_{nk}(\mathcal{X}^{'})\bar{R}_{nk}(\mathcal{X}^{'}\cup\{s\})}\mathbb{E}\left\{ B\log\left(1+\frac{|h_{km^{*}}|^{2}\mathrm{SINR}_{m^{*}}}{1+Y_{nk}(\mathcal{X}^{'})}\right)\right\} \\
\underset{>}{(b)} & \frac{\left|f_{n}\right|}{\bar{R}_{nk}(\mathcal{X}^{'})\bar{R}_{nk}(\mathcal{X}^{'}\cup\{s\})}\mathbb{E}\left\{ B\log\left(1+\frac{|h_{km^{*}}|^{2}\mathrm{SINR}_{m^{*}}}{1+Y_{nk}(\mathcal{X}^{'})}\right)\right\} \\
= & \tilde{D}_{nk}\left(\mathcal{X}^{'}\cup\left\{ s\right\} \right)-\tilde{D}_{nk}\left(\mathcal{X}^{'}\right),
\end{split}
\]
where the inequality (a) comes from $\bar{R}_{nk}(\mathcal{X})\leq\bar{R}_{nk}(\mathcal{X}^{'})$
and $\bar{R}_{nk}(\mathcal{X}\cup\{s\})\leq\bar{R}_{nk}(\mathcal{X}^{'}\cup\{s\})$,
and the inequality (b) holds since $Y_{nk}(\mathcal{X})<Y_{nk}(\mathcal{X}^{'})$
and $B\log\left(1+\frac{|h_{km^{*}}|^{2}\mathrm{SINR}_{m^{*}}}{1+Y_{nk}(\mathcal{X})}\right)>B\log\left(1+\frac{|h_{km^{*}}|^{2}\mathrm{SINR}_{m^{*}}}{1+Y_{nk}(\mathcal{X}^{'})}\right)$.

\textbf{Case III: } $\mathcal{X}\subseteq\mathcal{X}'\in\mathcal{I}$
and $0=\sum_{m\in\mathcal{A}_{k}}x_{nm}<\sum_{m\in\mathcal{A}_{k}}x_{nm}^{'}$

We have $\tilde{D}_{nk}\left(\mathcal{X}\cup\left\{ s\right\} \right)-\tilde{D}_{nk}\left(\mathcal{X}\right)=D_{nk}+\frac{\left|f_{n}\right|}{\bar{R}_{nk}(\mathcal{X}_{k})}-\frac{\left|f_{n}\right|}{\bar{R}_{nk}(\{s\})}$.
The following inequality 
\[
\begin{split} & \tilde{D}_{nk}\left(\mathcal{X}\cup\left\{ s\right\} \right)-\tilde{D}_{nk}\left(\mathcal{X}\right)=D_{nk}+\frac{\left|f_{n}\right|}{\bar{R}_{nk}(\mathcal{X}_{k})}-\frac{\left|f_{n}\right|}{\bar{R}_{nk}(\{s\})}\\
> & \frac{\left|f_{n}\right|}{\bar{R}_{nk}(\mathcal{X}^{'})}-\frac{\left|f_{n}\right|}{\bar{R}_{nk}\left(\mathcal{X}^{'}\cup\left\{ s\right\} \right)}=\tilde{D}_{nk}\left(\mathcal{X}^{'}\cup\left\{ s\right\} \right)-\tilde{D}_{nk}\left(\mathcal{X}^{'}\right)
\end{split}
\]
is satisfied, since $D_{nk}+\frac{\left|f_{n}\right|}{\bar{R}_{nk}(\mathcal{X}_{k})}>\frac{\left|f_{n}\right|}{\bar{R}_{nk}(\mathcal{X}^{'})}$
and $\frac{\left|f_{n}\right|}{\bar{R}_{nk}(\{s\})}<\frac{\left|f_{n}\right|}{\bar{R}_{nk}\left(\mathcal{X}^{'}\cup\left\{ s\right\} \right)}$.
In this case, we still get $\tilde{D}_{nk}\left(\mathcal{X}\cup\left\{ s\right\} \right)-\tilde{D}_{nk}\left(\mathcal{X}\right)>\tilde{D}_{nk}\left(\mathcal{X}^{'}\cup\left\{ s\right\} \right)-\tilde{D}_{nk}\left(\mathcal{X}^{'}\right)$.

 Combining the above three cases, we have 
\begin{equation}
\tilde{D}_{nk}\left(\mathcal{X}\cup\left\{ s\right\} \right)-\tilde{D}_{nk}\left(\mathcal{X}\right)\geq\tilde{D}_{nk}\left(\mathcal{X}^{'}\cup\left\{ s\right\} \right)-\tilde{D}_{nk}\left(\mathcal{X}^{'}\right).
\end{equation}
Meanwhile, it is trivial to show that since $\bar{R}_{nk}(\mathcal{X})\leq\bar{R}_{nk}(\mathcal{X}^{'})$,
we have $\tilde{D}_{nk}\left(\mathcal{X}\right)\leq\tilde{D}_{nk}\left(\mathcal{X}'\right)$
for any $\mathcal{X}\subseteq\mathcal{X}'$. Therefore, $\tilde{D}_{nk}\left(\mathcal{X}\right)$
is a monotone submodular function. In the above discussion, cooperative
beamforming is applied as a candidate transmission scheme to demonstrate
the monotone submodular property of the average delay function. In
fact, this property holds for any candidate transmission scheme.

\subsection{Basics of the Message Passing Procedure\label{subsec:message_passing} }

We briefly introduce the factor graph model and the max-product algorithm.
A factor graph is a bipartite graph which consists of $I$ variable
nodes $\{\mu_{1},\cdots,\mu_{I}\}$ and $J$ function nodes $\{F_{1},\cdots,F_{J}\}$.
Let $\Gamma_{i}^{\mu}$ and $\Gamma_{j}^{F}$ denote the set of indices
of the neighboring function nodes of a variable node $\mu_{i}$ and
that of the neighboring variable nodes of a function node $F_{j}$,
respectively. Max-product is a belief propagation algorithm based
on the factor graph model, which is widely applied to find the optimum
of the global function taking the form as $F(\bm{\mu})=\prod_{j=1}^{J}F_{j}(\mu_{\varGamma_{j}^{F}})$
in a distributed manner. A comprehensive tutorial can be found in
\cite{kschischang_factor_2001}. 

In each iteration, each variable node sends one updated message to
one of its neighboring function nodes and receives one updated message
from this node. According to the max-product algorithm \cite{kschischang_factor_2001},
the message from a variable node $\mu_{i}$ to a function node $F_{j}$,
i.e., $m_{\mu_{i}\rightarrow F_{j}}^{t}(x)$, is updated as 
\begin{equation}
m_{\mu_{i}\rightarrow F_{j}}^{t+1}(x)=\prod_{l\in\Gamma_{i}^{\mu}\backslash\{j\}}m_{F_{l}\rightarrow\mu_{i}}^{t}(x),\label{eq:variable-function}
\end{equation}
 which collects all the beliefs on the value of $\mu_{i}=x$ from
the neighboring function nodes $F_{l}$ $(l\in\Gamma_{i}^{\mu}\backslash\{j\})$
except $F_{j}$. The message from a function node $F_{j}$ to a variable
node $\mu_{i}$, i.e., $m_{F_{j}\rightarrow\mu_{i}}^{t}(x)$, is updated
as
\begin{equation}
m_{F_{j}\rightarrow\mu_{i}}^{t+1}(x)=\max_{\Gamma_{j}^{F}\backslash\{i\}}\left\{ F_{j}(\bm{X})\prod_{l}m_{\mu_{l}\rightarrow F_{j}}^{t}(x_{l})\right\} ,\label{eq:function-variable}
\end{equation}
which achieves the maximization of the product of the local function
$F_{j}(\bm{X})$ and incident messages over configurations in $\Gamma_{j}^{F}\backslash\{i\}$.

\subsection{\label{subsec:Proof-of-Theorem-message-update}Proof of Theorem $\ref{thm:message_update}$}

By substituting (\ref{eq:message_alpha}) into (\ref{eq:variable-function}),
we can easily obtain the practical message $\alpha_{i\rightarrow j}^{t}$
as given by (\ref{eq:alpha_variable_function_ratio}).

From (\ref{eq:function-variable}), the derivation of the message
$\beta_{j\rightarrow i}^{t}$ involves one maximization operation
over all possible values of $\{\mu_{l}=x_{l}\}$ $(l\in\Gamma_{j}^{F}\backslash\{i\})$.
Then, we discuss the message $\beta_{j\rightarrow i}^{t}$ in the
cases when $F_{j}\doteq\eta_{nk}$ and $F_{j}\doteq g_{m}$, respectively.

\textbf{Case I: } Derivation of $\beta_{j\rightarrow i}^{t}$ for
$F_{j}\doteq\eta_{nk}$

By substituting the average delay (such as the metric presented in
(\ref{eq:av_Delay})) into (\ref{eq:function-variable}), the message
$m_{F_{j}\rightarrow\mu_{i}}^{t+1}(1)$ with $F_{j}=\eta_{nk}$ and
$\mu_{i}=1$ can be represented as\begin{small}
\begin{equation}
\begin{split}m_{F_{j}\rightarrow\mu_{i}}^{t+1}(1)= & \max_{E_{i}^{1}}\left\{ \exp(-p_{nk}\bar{D}_{nk}(\bm{X}^{(1)}))\prod_{l\in E_{i}^{1}}\left(\frac{m_{\mu_{l}\rightarrow F_{j}}^{t}(1)}{m_{\mu_{l}\rightarrow F_{j}}^{t}(0)}\right)\right\} \\
 & \times\prod_{l\in\Gamma_{j}^{F}\backslash\{i\}}m_{\mu_{l}\rightarrow F_{j}}^{t}(0),
\end{split}
\label{eq:function-variable-xe-one}
\end{equation}
\end{small}where $E_{i}^{1}\subseteq\Gamma_{j}^{F}\backslash\{i\}$
is a subset of the index set $\Gamma_{j}^{F}\backslash\{i\}$ such
that its associated elements in $\bm{X}^{(1)}$ are equal to one,
$i.e.$, $\mu_{l}=1$ for all $l\in E_{i}^{1}\cup\{i\}$, while $\mu_{l}=0$
for all $l\in\Gamma_{j}^{F}\backslash\{i\}\backslash E_{i}^{1}$.
Similarly, we can compute the message $m_{F_{j}\rightarrow\mu_{i}}^{t+1}(0)$
as \begin{small}

\begin{equation}
\begin{split}m_{F_{j}\rightarrow\mu_{i}}^{t+1}(0)= & \max_{E_{i}^{2}}\left\{ \exp(-p_{nk}\bar{D}_{nk}(\bm{X}^{(0)}))\prod_{l\in E_{i}^{2}}\left(\frac{m_{\mu_{l}\rightarrow F_{j}}^{t}(1)}{m_{\mu_{l}\rightarrow F_{j}}^{t}(0)}\right)\right\} \\
 & \times\prod_{l\in\Gamma_{j}^{F}\backslash\{i\}}m_{\mu_{l}\rightarrow F_{j}}^{t}(0)
\end{split}
\label{eq:function-variable-xe-zero}
\end{equation}
\end{small}where $E_{i}^{2}\subseteq\Gamma_{j}^{F}\backslash\{i\}$
is also a subset of the index set $\Gamma_{j}^{F}\backslash\{i\}$
such that its associated elements in $\bm{X}^{(0)}$ are equal to
one, while the other elements are zero with $\mu_{l}=0$ for all $l\in\Gamma_{j}^{F}\backslash E_{i}^{2}$.
From (\ref{eq:function-variable-xe-one}) and (\ref{eq:function-variable-xe-zero}),
the message  $\beta_{j\rightarrow i}^{t+1}$ can be expressed as\begin{small}
\begin{equation}
\begin{split}\beta_{j\rightarrow i}^{t+1} & =\max_{E_{i}^{1}}\left\{ (-p_{nk}\bar{D}_{nk}(\bm{X}^{(1)}))+\sum_{l\in E_{i}^{1}}\alpha_{l\rightarrow j}^{t}\right\} \\
 & \,\,-\max_{E_{i}^{2}}\left\{ (-p_{nk}\bar{D}_{nk}(\bm{X}^{(0)}))+\sum_{l\in E_{i}^{2}}\alpha_{l\rightarrow j}^{t}\right\} ,\\
 & =p_{nk}\left(\bar{D}_{nk}(\bm{X}_{i}^{(0)})-\bar{D}_{nk}(\bm{X}_{i}^{(1)})\right),
\end{split}
\label{eq:beta_function_variable_ratio_v2}
\end{equation}
\end{small}where $\bm{X}_{i}^{(0)}$ and $\bm{X}_{i}^{(1)}$ are
set as caching vectors by selecting the variable nodes $\{\mu_{l}\}$
with positive $\alpha_{l\rightarrow j}^{t}$, $i.e.$, $l\in E_{i}^{+}=\{i^{'}\in\Gamma_{j}^{F}\backslash\{i\}|\alpha_{i^{'}\rightarrow j}^{t}>0\}$,
and assigning their associated elements to one. Thus, we have $\mu_{l}\doteq x_{nm}=1$
for all $l\in E_{i}^{+}$ in $\bm{X}_{i}^{(0)}$ and $\mu_{l}\doteq x_{nm}=1$
for all $l\in E_{i}^{+}\cup\{i\}$ in $\bm{X}_{i}^{(1)}$.  This
means that each function node $F_{j}$ should select its neighboring
variable nodes $\mu_{l}$ with positive input message $\alpha_{l\rightarrow j}^{t}$
and then calculate the delay gap between $\bar{D}_{nk}(\bm{X}_{i}^{(0)})$
and $\bar{D}_{nk}(\bm{X}_{i}^{(1)})$. 

\textbf{Case II: } Derivation of $\beta_{j\rightarrow i}^{t}$ for
$F_{j}\doteq g_{m}$ 

By substituting the constraint function into (\ref{eq:function-variable}),
the message $m_{F_{j}\rightarrow\mu_{i}}^{t+1}(1)$ when $F_{j}\doteq g_{m}$
can be represented as\begin{small} 
\begin{equation}
\begin{split}m_{F_{j}\rightarrow\mu_{i}}^{t+1}(1)= & \max_{E_{i}^{3}}\left\{ g_{m}(\bm{X}^{(1)})\prod_{l\in E_{i}^{3}}\left(\frac{m_{\mu_{l}\rightarrow F_{j}}^{t}(1)}{m_{\mu_{l}\rightarrow F_{j}}^{t}(0)}\right)\right\} \\
 & \times\prod_{l\in\Gamma_{j}^{F}\backslash\{i\}}m_{\mu_{l}\rightarrow F_{j}}^{t}(0),
\end{split}
\label{eq:function-variable-xe-one-1}
\end{equation}
\end{small}where $E_{i}^{3}$ is a subset of the index set $\Gamma_{j}^{F}\backslash\{i\}$
and $|E_{i}^{3}|\leq Q_{m}-1$. This means that to satisfy the cache
capacity constraint, there exist at most $Q_{m}-1$ neighboring variable
nodes $\{\mu_{l}\}$ with $\mu_{l}=1$ $(l\in E_{i}^{3})$ except
the variable node $\mu_{i}=1$. Similarly, we can compute the message
$m_{F_{j}\rightarrow\mu_{i}}^{t+1}(0)$ when $F_{j}\doteq g_{m}$
as\begin{small} 
\begin{equation}
\begin{split}m_{F_{j}\rightarrow\mu_{i}}^{t+1}(0)= & \max_{E_{i}^{4}}\left\{ g_{m}(\bm{x}^{(0)})\prod_{l\in E_{i}^{4}}\left(\frac{m_{\mu_{l}\rightarrow F_{j}}^{t}(1)}{m_{\mu_{l}\rightarrow F_{j}}^{t}(0)}\right)\right\} \\
 & \times\prod_{l\in\Gamma_{j}^{F}\backslash\{i\}}m_{\mu_{l}\rightarrow F_{j}}^{t}(0)
\end{split}
\label{eq:function-variable-xe-zero-1}
\end{equation}
\end{small}where $E_{i}^{4}$ is a subset of the index set $\Gamma_{j}^{F}\backslash\{i\}$
and $|E_{i}^{4}|\leq Q_{m}$. Since $\mu_{i}=0$, there exist at most
$Q_{m}$ neighboring variable nodes $\{\mu_{l}\}$ $(l\in E_{i}^{4})$
with $\mu_{l}=1$ to satisfy the cache capacity constraint. From (\ref{eq:function-variable-xe-one-1})
and (\ref{eq:function-variable-xe-zero-1}), the message ratio of
$m_{F_{j}\rightarrow\mu_{i}}^{t+1}(1)$ and $m_{F_{j}\rightarrow\mu_{i}}^{t+1}(0)$
in the logarithmic domain can be expressed as\begin{small}
\begin{equation}
\begin{split}\beta_{j\rightarrow i}^{t+1} & =\max_{E_{i}^{3}}\left\{ \sum_{l\in E_{i}^{3}}\alpha_{l\rightarrow j}^{t}\right\} -\max_{E_{i}^{4}}\left\{ \sum_{l\in E_{i}^{4}}\alpha_{l\rightarrow j}^{t}\right\} \end{split}
.\label{eq:log-nodes-edges-v}
\end{equation}
\end{small}By sorting the messages $\{\alpha_{l\rightarrow j}^{t}\}$
($\forall l\in\Gamma_{j}^{F}\backslash\{i\}$) in the decreasing order
as $\alpha_{l\rightarrow j}^{(1)}$, $\alpha_{l\rightarrow j}^{(2)}$,
$\cdots$, $\alpha_{l\rightarrow j}^{(Q_{m}-1)}$, $\cdots$ , we
can further simplify $\beta_{j\rightarrow i}^{t+1}$ as\begin{small}
\begin{equation}
\begin{split}\beta_{j\rightarrow i}^{t+1} & =\begin{cases}
\min\{0,-\alpha_{l\rightarrow j}^{(Q_{m})}\}, & \text{if }\alpha_{l\rightarrow j}^{(Q_{m}-1)}\geq0,\\
0, & \text{otherwise},
\end{cases}\end{split}
\label{eq:xi_function_variable_ratio_v2}
\end{equation}
\end{small}which is exactly equal to $\min\{0,-\alpha_{l\rightarrow j}^{(Q_{m})}\}$,
as given by (\ref{eq:xi_function_variable_ratio}).

\bibliographystyle{IEEEtran}

\end{document}